@arxiver{icm_hr_nt_1000.pdf,profile_hr_nt.pdf,bub_hr_t_1000.pdf}

\documentclass[fleqn,usenatbib]{mnras}

\usepackage{newtxtext,newtxmath}

\usepackage[T1]{fontenc}
\usepackage{ae,aecompl}

\usepackage{graphicx}	
\usepackage{amsmath}	
\usepackage{amssymb}	
\usepackage{geometry}
\usepackage{float}
\usepackage{subcaption}
\usepackage{hyperref}

\usepackage{eucal}

\newcommand{\ramses}{\textsc{ramses}}
\newcommand{\gizmo}{\textsc{gizmo}}
\newcommand{\sub}[1]{_{\mathrm{#1}}}
\newcommand{\rhoa}{\rho\sub{amb}}
\newcommand{\rhob}{\rho\sub{bub}}
\newcommand{\msun}{M$_{\sun}$}

\def\equationautorefname~#1\null{Eq.~(#1)\null}
\def\figureautorefname~#1\null{Fig.~#1\null}
\newcommand{\appref}[1]{\hyperref[#1]{Appendix~\ref{#1}}}

\defcitealias{Komatsu2001}{KS01}

\title[Stability of AGN bubbles]{Physical and numerical stability and instability of AGN bubbles in a hot intracluster medium}
\author[Ogiya et al.]{
Go Ogiya$^{1}$\thanks{E-mail: go.ogiya@oca.eu (GO)}, 
Pawel Biernacki$^{2}$\thanks{E-mail: biernack@physik.uzh.ch (PB)},
Oliver Hahn$^{1}$
and Romain Teyssier$^{2}$
\\
$^{1}$Laboratoire Lagrange, Universit\'e C\^ote d'Azur, Observatoire de la C\^ote d'Azur, CNRS,\\ \quad Blvd de l'Observatoire, CS 34229, 06304 Nice, France\\
$^{2}$Center for Theoretical Astrophysics and Cosmology, Institute for Computational Science,\\ \quad University of Zurich, Winterthurerstrasse 190, 8057 Zurich, Switzerland
}

\date{Accepted XXX. Received YYY; in original form ZZZ}

\pubyear{2018}

\begin{document}
\label{firstpage}
\pagerange{\pageref{firstpage}--\pageref{lastpage}}
\maketitle

\begin{abstract} 
While feedback from Active Galactic Nuclei (AGN) is an important heating source in the centre of galaxy clusters, it is still unclear how the feedback energy is injected into the intracluster medium (ICM) and what role different numerical approaches play. 
Here, we compare four hydrodynamical schemes in idealized simulations of a rising bubble inflated by AGN feedback in a hot stratified ICM: (traditional) smoothed particle hydrodynamics (TSPH), a pressure flavour of SPH (PSPH), a meshless finite mass (MFM) scheme, as well as an Eulerian code with adaptive mesh refinement. In the absence of magnetic fields, the bubble is Kelvin-Helmholtz unstable on short enough time scales to dissolve it fully in the ICM, which is captured by MFM and \ramses{} simulations, while in the TSPH simulation the bubble survives.
When the ICM is turbulent, mixing of the bubble with the ICM is accelerated. This occurs if the numerical scheme can capture the instabilities well. The differences in the evolution of the bubble has a surprisingly small influence on the thermal structure of the ICM. However, in the simulations with MFM and \ramses{} the bubble disruption leads to turbulent stirring of the ICM which is suppressed in SPH. In the latter the thermal energy remains trapped in the bubble and is transported to large radii. We discuss if the choice of hydrodynamical schemes can lead to systematic differences in the outcomes of cosmological simulations.
\end{abstract}

\begin{keywords}
methods: numerical -- galaxies: clusters: intracluster medium -- hydrodynamics -- instabilities -- turbulence
\end{keywords}



\section{Introduction} \label{sec:intro}
The heating mechanisms for the gas in the centre of galaxy clusters is important to explain X-ray observations \citep[][and references therein]{McNamara2007, Kravtsov2012, Fabian2012}. The cooling timescale of the gas in the centres of galaxy clusters is much shorter than the Hubble time \cite[][and many subsequent works]{Fabian1977, Cowie1977, Mathews1978}, known as the so-called cooling flow problem. If there is no heating source, intra-cluster medium (ICM) would exhibit a strong cooling flow and thus become highly concentrated in the centre of the cluster. This is in contrast with observations of cluster centres, which show that the ratio of the gas mass to the total enclosed mass can be fairly low, $\sim 0.01$, compared to the cosmic baryon fraction, $\sim 0.16$. Furthermore, the temperature of the central gas is high, typically $\ga {\rm keV} \sim 10^7$\,K \citep[e.g.][]{Lea1973, Evrard1997, Pratt2010, Mantz2014}. These observed properties of the ICM are reviewed by e.g. \cite{Fabian1994} and \citet[][and references therein]{Fabian2012}. 

Active Galactic Nuclei (AGN) are promising heating sources to keep the centres of galaxy clusters hot and to prevent gas concentration \citep[][and references therein]{McNamara2007, Kravtsov2012}. If AGN are powerful enough, the jet power of $P\sub{jet} \ga 10^{42}$ erg/sec \citep[e.g.][]{Cavagnolo2010, Nemmen2012, Godfrey2013}, can trigger strong shock waves which compress and heat the ICM \citep[e.g.][]{Sanderson2005, Sutherland2007, Mingo2012, Gaspari2011, Wagner2012, Perucho2014, Lanz2015}. Even if AGN are less powerful, they may induce bubbles of diffuse hot gas. X-ray observations have detected such AGN bubbles as cavities in the ICM \citep{Fabian2000, McNamara2001, Gitti2006, Dong2010} and in galaxy groups and galaxies \citep[e.g.][]{Ohto2003, Forman2007, Panagoulia2014}. 

Observations have also suggested that significant amounts of the thermal energy might be still captured in hot bubbles \citep[e.g.][]{Birzan2004, Dunn2005, Shurkin2008, Sanders2009}. Hence, a theoretical investigation of the interaction between the bubble and the surrounding ambient cluster gas would be of great importance to our understanding of the thermodynamics of the ICM. The rising bubble may also play a role in redistributing heavy elements in the ICM. The processes that govern the rising of the bubble and its subsequent mixing with the surrounding ICM are complex and detailed analytical investigations are difficult. Simplified models have provided valuable insights \citep[e.g.][and references therein]{Voit2017}, and numerical simulations can provide further insight into the complex processes governing gas in cluster cores.

Numerical simulations of the ICM with AGN feedback can be classified into two types - 1) idealised and 2) cosmological. The former aim to understand the physics by means of idealised setups, which provide full control over the cluster environment and processes. For example, early studies showed that energy and matter redistribution by rising bubbles plays a key role in solving the cooling flow problem \citep[e.g.][]{Churazov2001, Bruggen2002}. Subsequent simulations with higher resolution and additional physics, including magnetic fields and cosmic rays, demonstrated that the buoyantly rising bubbles redistribute not only energy and mass, but also metals and magnetic fields in the ICM \citep{Reynolds2005, Sijacki2006, Vernaleo2006, Roediger2007, Dursi2008, Vazza2010, Guo2011}. An interesting insight obtained by magnetohydrodynamic (MHD) simulations is that magnetic tension suppresses mixing instabilities on the bubble surface and thus supports bubbles to rise to larger radii \citep[][see also Biernacki et al., in prep.]{Robinson2004, Dong2009}. 

Cosmological simulations provide a more realistic cluster environment and assembly history with turbulence in the ICM driven continuously by both minor and major mergers \citep[e.g.][]{Miniati2014}. Thanks to recent developments in the modelling of supermassive black holes and AGN feedback in cosmological hydrodynamic simulations \citep[e.g.][]{DiMatteo2005, Kawata2005, Sijacki2007, Okamoto2008, Booth2009, Teyssier2011, Vogelsberger2013, Steinborn2015}, they have succeeded in reproducing various observational results \citep[][and refereces therein]{Nagai2007, McCarthy2010, DiMatteo2012, Battaglia2013, LeBrun2014, Dolag2016, Dubois2016, Barnes2017_IllustrisTNG, Barnes2017_ClusterEAGLE, Hahn2017}. 

However, full agreement among them has not been achieved yet. The early discrepancy in predicted cluster entropy profiles between Eulerian and Lagrangian methods in non-radiative simulations \citep{Frenk1999} has been understood as a severe underproduction of entropy in traditional Smooth Particle Hydrodynamics (SPH) methods \citep[e.g.][]{Wadsley2008, Mitchell2009, Power2014}. Modern Lagrangian methods solved these shortcomings, but such discrepancies appear in any case less dramatic when optically thin cooling is added and are overshadowed by differences in subgrid models \citep[e.g.][]{Sembolini2016}. However, some suspicion about fundamental differences might still be in order: for example, the central gas mass fraction in the centres of simulated clusters is typically lower than that observed when grid-based hydrodynamical solvers are used in the simulations \citep[][]{Hahn2017, Barnes2017_IllustrisTNG}. On the other hand, simulations that adopt SPH claim to reproduce more realistic gas fractions by carefully tuning the AGN feedback model \citep{Battaglia2013,LeBrun2014}. Similarly, the distribution of metals in galaxy clusters is reproduced in some SPH simulations \citep{Wiersma2011,Planelles2014,Rasia2015}.
\cite{Schaller2015} found differences in simulations adopting different flavours of SPH when keeping the feedback models fixed, but the differences are small at galaxy cluster masses.

Such discrepancies motivated us to pose the following questions: 
\begin{itemize}
\item are there issues in AGN feedback modelling?
\item do different hydrodynamical solvers agree? 
\item do the simulations lack the resolution to capture important processes? 
\item are we missing any of other important non-thermal processes?
\end{itemize}

Physical viscosity and diffusion are typically negligible in most processes of the formation and evolution of galaxy clusters \citep[e.g.][]{Mo2010}. However, neglecting this physics in numerical simulations may lead the code-dependent numerical, i.e. artificial and physically incorrect, effects which can affect the outcomes of the simulations. Thus we need to give careful attention to this point. 
\cite{Agertz2007} presented the fundamental differences between SPH and grid based methods with a suite of idealised simulations of a cold dense gas cloud moving through a low-density hot medium.
\cite{Wadsley2008} tackled the second question using idealized simulations of a buoyantly unstable and rising hot bubble in an ambient medium. They found that the absence of mixing in traditional SPH schemes leads to an underproduction of entropy compared to grid based codes.

In this paper, we update the findings by \cite{Agertz2007} and \cite{Wadsley2008} using state-of-the-art hydrodynamical solvers that are used in more recent major cosmological simulations. In order to avoid too much complexity, we employ a well-defined, simple setup of a spherically symmetric ICM with a hot bubble inflated by AGN feedback. This setup is similar to the one of \citet{Wadsley2008}, but includes self-gravity of the gas, and for which a na\"ive analytical expectation can be given. In a next step, we introduce a turbulent velocity field to make the model more realistic. We address also the third question by varying the resolution of the simulations. For the first question, we refer readers to \cite{Meece2017} who compared the commonly-adopted sub-grid models of AGN feedback. Regarding the fourth question, our subsequent project will address one of the possibilities - effects of magnetic fields (Biernacki et al., in prep.).

This paper is organized as follows. In \autoref{sec:analytical_exp}, we describe our simple model of a buoyantly rising bubble inflated by AGN feedback at the centre of a gas sphere of the ICM and also provide the analytical expectation for the fate of the bubble. \autoref{sec:hydro_codes} gives a brief description for the numerical codes and methods used in this paper. We describe the setup of our numerical experiments and demonstrate the results in \autoref{sec:sims}. In \autoref{sec:summary_discussion}, we summarize and discuss the results.

\section{Initial Conditions and Analytic Expectations} 
\label{sec:analytical_exp}

In this section, we set the stage for our model of a stratified hydrostatic medium and describe a hot bubble positioned initially near the centre of our idealised cluster. We present also calculations that demonstrate that such bubbles will buoyantly rise, experience ram pressure and undergo interface instabilities that lead to their ultimate demise by mixing with the ambient medium.  

\subsection{The ambient medium} \label{ssec:KS_model}

We adopt the analytical model proposed by \citet[hereafter \citetalias{Komatsu2001}]{Komatsu2001} as a model for the ambient medium. In the analytical model, a gas sphere is embedded in a dark matter (DM) halo with a Navarro-Frenk-White \citep[NFW][]{Navarro1997} density profile, and the thermal pressure balances with the gravity of the DM halo with a polytropic equation of state while the self-gravity of the gas is neglected. \citetalias{Komatsu2001} adopted empirical prescriptions to give the concentration parameter of the halo, $c = r\sub{100}/r\sub{s}$, where $r\sub{s}$ is the scale length of the halo, and the polytropic index, $\gamma$, as functions of the virial mass of the DM halo, $M\sub{100}$. Here, $M\sub{100}$ is the mass contained within the virial radius, $r\sub{100}$, inside of which, the mean density of the DM halo is 100 times the critical density of the current universe. 

Throughout this paper, we adopt a Hubble constant $H_0 = 70.3$\,{km/s/Mpc} \citep{Komatsu2011}  and we assume $M\sub{100} = 3\times10^{14}$\,\msun{} and a total gas mass, $M\sub{gas} = 4.5\times10^{13}$\,\msun{}. The concentration parameter of the DM halo and effective polytropic exponent that the analytical model provides are $c = 5.168$ and $\gamma = 1.137$, respectively. The DM halo has a virial radius of $r\sub{100} = 1.734$\,Mpc. The dashed line in \autoref{fig:nofeedback_1000} presents the radial profiles of gas density (first row) and temperature (second row) given by this model.

\subsection{Equation of motion of a bubble} \label{ssec:eom_bub}

\subsubsection{Buoyancy}
Let us consider a hot underdense bubble embedded in a colder stratified ambient gas sphere in hydrostatic equilibrium. If the bubble is displaced from the centre of the ambient gas sphere, it will rise buoyantly if the Schwarzschild condition for convective stability \citep{Schwarzschild1906}, 
\begin{eqnarray}
\biggl | \frac{dT}{dr} \biggr | < \biggl | \frac{dT}{dr} \biggr |\sub{ad}
\label{eq:stab_cond}
\end{eqnarray}
is {\em not fulfilled}. 
Here, $r$ and $T$ are the distance from the centre of the gas sphere and the gas temperature, respectively. The subscript ``ad'' in the second term indicates the respective relation allowing only adiabatic processes. Treating the bubble as a point mass for simplicity, the acceleration due to buoyancy, ${\bf a}\sub{buo}$, can be written as 
\begin{eqnarray}
{\bf a}\sub{buo}({\bf r}) = \frac{\rhoa(r) - \rhob(r)}{\rhob(r)} {\bf g}({\bf r}), \label{eq:buoyancy}
\end{eqnarray}
where ${\bf r}$ is the position of the bubble relative to the centre, ${\bf g}({\bf r})$ the gravitational acceleration at ${\bf r}$, and $\rhoa(r)$ and $\rhob$ indicate the density of the ambient medium at the radius, $r = |{\bf r}|$, and of the bubble, respectively. Note that the first term equates with the gradient of the thermal pressure, $-dp/dr$. 

\subsubsection{Ram Pressure}
When the bubble has a non-zero velocity with respect to the ambient medium, it also feels ram pressure. Treating the bubble again as a point mass, the gradient of ram pressure, ${\bf a}\sub{ram}$, is given as
\begin{eqnarray}
{\bf a}\sub{ram}({\bf r}, v) = \frac{v^2}{2 \rhob} \frac{d\rhoa(r)}{dr} \frac{{\bf r}}{r}, \label{eq:ram_pressure}
\end{eqnarray}
where $v$ is the relative velocity between the bubble and the ambient medium.

\subsubsection{Interface instabilities} \label{ssec:KHI}
As the bubble moves through the ambient medium with non-zero $v$, we expect the bubble boundary to undergo a Kelvin-Helmholtz instability \citep[KHI, cf. e.g.][]{Landau1959} which ultimately will act to dissolve the bubble in the ambient medium. The timescale on which the KHI will act to dissolve the bubble is 
\begin{eqnarray}
\tau\sub{KHI} \sim \frac{\rhoa + \rhob}{\sqrt{\rhoa \rhob}} \frac{\lambda}{v}, \label{eq:tau_khi}
\end{eqnarray}
where $\lambda$ is the wavelength of the perturbation on the surface of the bubble. The KHI grows exponentially with time, $t$, i.e., the amplitude of the KHI, $A\sub{KHI} \propto \exp{(t/\tau\sub{KHI})}$.

\subsection{Comparison of timescales} \label{ssec:t_comp}
\begin{figure}
\begin{center}
\includegraphics[width=0.8\columnwidth]{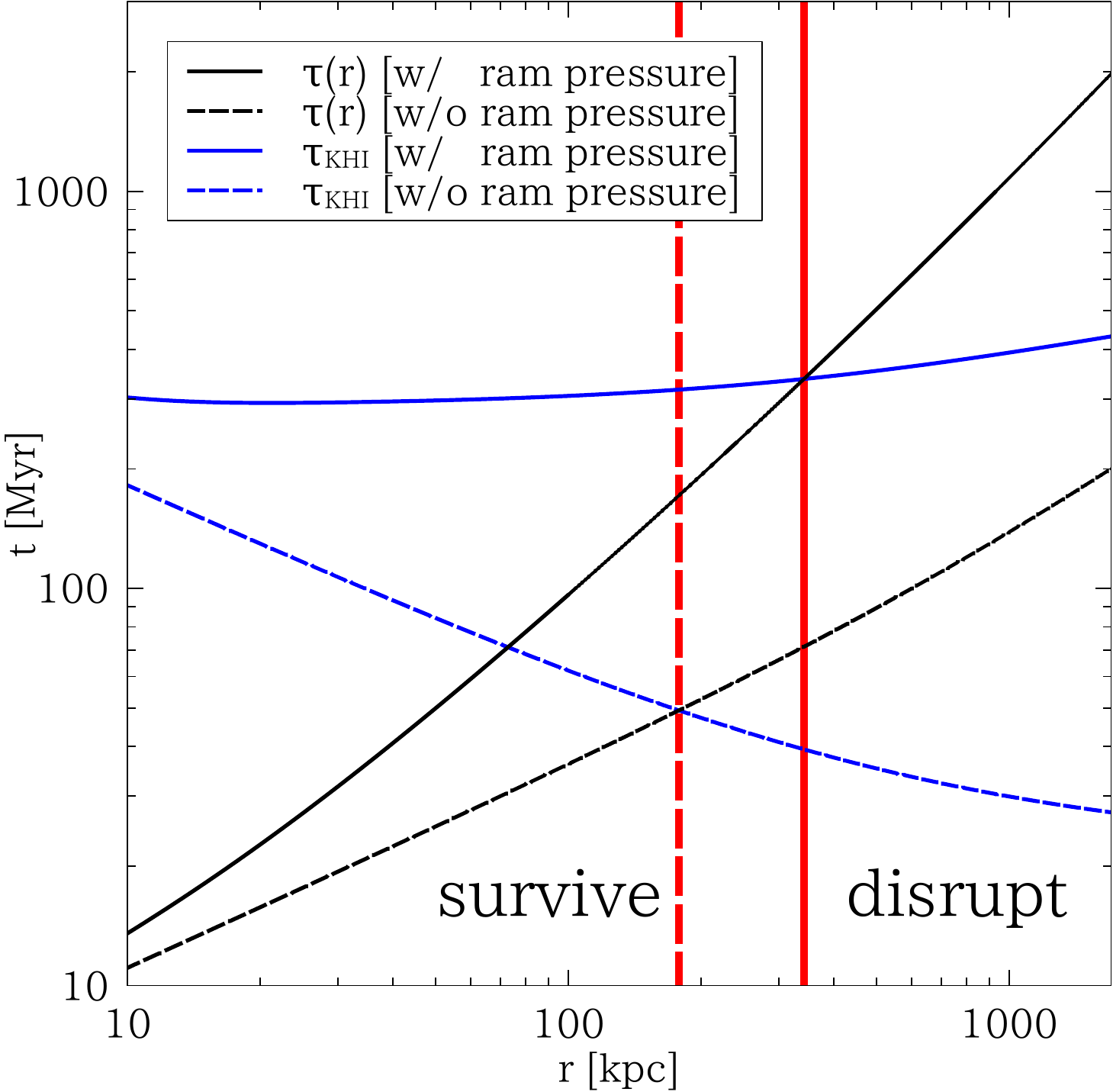}
\end{center}
\caption{
Comparison of timescales. Black lines represent the timescale of the bubble to arrive at $r$, $\tau(r)$. To estimate the KHI timescale, $\tau\sub{KHI}$ (blue), the instantanious relative velocity between the bubble and the ambient medium at $r$, $v(r)$, and the wavelength, $\lambda = 59.3$\,kpc are assumed. For solid (dashed) lines, ram pressure is (is not) taken into account. The bubble is expected to be dissolved at $r \sim 200-300$\,kpc where $\tau(r) > \tau\sub{KHI}$. Red solid and dashed lines indicate the survival radius of the bubble including and not including ram pressure. 
\label{fig:tbuoyancy_vs_tkhi}}
\end{figure}

Based on our estimates above, the bubble is expected to survive until $\lesssim \tau\sub{KHI}$ but dissolve after that. Using \autoref{eq:buoyancy} and \autoref{eq:ram_pressure}, we can follow the motion of a bubble and estimate the time for the bubble to arrive at $r$, $\tau(r)$. To make a hot and less dense bubble, the temperature of gas contained within the innermost sphere of a diameter of $20$\,kpc is set to be $10^9$\,K, much hotter than the ambient hydrostatic medium (see \autoref{fig:nofeedback_1000}). Assuming that the heated central sphere itself expands adiabatically behind a rapidly propagating shock (see also \autoref{fig:bub_cen_100_profiles}) until its pressure balances that of the ambient medium, the diameter of the bubble becomes $\sim60$\,kpc (with the assumption of adiabatic expansion of the bubble rendering this an upper limit).
If the position of the bubble is displaced from the exact centre by some perturbation, it rises buoyantly as long as \autoref{eq:stab_cond} is not satisfied. 

\autoref{fig:tbuoyancy_vs_tkhi} compares $\tau(r)$ with $\tau\sub{KHI}$ and shows that the bubble would be dissolved at $r \ga 200-300$\,kpc where $\tau(r) > \tau\sub{KHI}$. We compute the bubble density by assuming that the bubble continues to expand adiabatically and use the instant relative velocity between the bubble and the ambient medium at $r$, $v(r)$, and the wavelength, $\lambda = 59.3$\,kpc to evaluate $\tau\sub{KHI}$. Since $\tau\sub{KHI}$ is proportional to the wavelength of the perturbation, this choice of $\lambda$ would provide the upper limit of the of the KHI-time. Note that we neglect the response of the ambient medium to the motion and expansion of the bubble and assume that its density and temperature stay those of the hydrostatic equilibrium state, for simplicity.  

\subsection{Introducing turbulence} 
\label{ssec:int_turb}
In contrast to our simplistic model above, the ICM is turbulent \citep[e.g.][and other observational results]{Schuecker2004,Hitomi2016} and a substantial amount of turbulent energy may be converted into thermal energy to heat the centre of clusters \citep[e.g.][]{Dennis2005,Zhuravleva2014,Eckert2017}. In fact, mass estimates of galaxy clusters via the thermal Sunyaev-Zel'dovich effect and assuming hydrostatic equilibrium appear inconsistent with both independent measures of galaxy cluster abundance \citep[cf.][]{Planck2016a} and independent mass estimates \citep{Hurier2017}, pointing possibly to a much larger contribution of non-thermal pressure than what is commonly found in simulations \citep[see also e.g.][]{Lau2009,Nelson2014}. 

Here, we introduce turbulence in order to quantify the additional effect it can have on the mixing of rising AGN bubbles in purely hydrostatic simulations, as well as the opposite effect, how the bubble can drive turbulence itself. We adopt a simple model in which we sample an isotropic Gaussian Kolmogorov velocity spectrum \citep[e.g.][]{Landau1959}, for which 
\begin{equation}
\left<\tilde{\mathbf{v}}(\mathbf{k})\cdot \tilde{\mathbf{v}}^\ast(\mathbf{k'})\right> \propto |k|^{-11/3}\delta_{D}(\mathbf{k}-\mathbf{k}'),
\end{equation}
where a tilde indicates a Fourier transformed field. An arbitrary velocity field can be decomposed into longitudinal and transversal modes, with the respective scalar potential $\phi_v$ and the vector potential $\mathbf{A}_v$, so that  
\begin{equation}
\mathbf{v} = \boldsymbol{\nabla}\phi_v + \boldsymbol{\nabla}\times\mathbf{A}_v.
\end{equation}
We assume here that the potentials are being given through multiplication of a random scalar field $G\{0,1\}$ and an independent random vector field $\mathbf{G}\{0,1\}$ (both of which are assumed to be fields of Gaussian white noise with zero mean and unit variance) with the square-root of the spectrum, so that in Fourier space
\begin{equation}
\tilde{\phi}_v(\mathbf{k}) = \alpha \phi_0\, k^{-17/6}\, \tilde{G}\{0,1\},\quad \tilde{\mathbf{A}}_v(\mathbf{k}) = \beta \phi_0\, k^{-17/6}\, \tilde{\mathbf{G}}\{0,1\}. 
\end{equation}
Here, $\phi\sub{0}$ is a normalisation constant, and the parameters $\alpha$ and $\beta$ can be used to adjust the relative importance of longitudinal and transversal velocity modes ($\alpha^2+\beta^2=1$ to maintain normalisation and we set $\alpha^2=\beta^2=0.5$). Note that the Kolmogorov spectrum typically holds only between the driving scale and the dissipation scale. In our model, we set the driving scale $k\sub{0}$ by hand and zero all modes with $k<k_0$ where $k\sub{0} = 2\pi/(r\sub{100}/4)$ in this paper, while we assume that the dissipation scale is unresolved. In practice, we create a random realisation of a velocity field on a mesh of resolution $N\sub{v}^3$, with $N\sub{v}=256$. The finite resolution automatically introduces a small-scale cut off of $2\pi/(r\sub{100}/128)$, and we assume that the one-point variance on the grid, $\sigma\sub{v}^2$ can be equated to a non-thermal temperature of $T\sub{nt}$. For the simulation codes used in this study, we add the respective velocity by interpolating from the particle/AMR cell positions to the grid on which we made a realisation of the velocity field. Then the corresponding thermal energy of $k\sub{B} T\sub{nt}$, is subtracted from each particle/AMR cell. This keeps the total energy of the system constant down to the level of Poisson noise in the grid and particle distribution.

\section{Methods} 
\label{sec:hydro_codes}
In order to compare theoretical expectations outlined in the previous section with full non-linear calculations, we run idealized hydrodynamical simulations of self-gravitating gas using two independently developed numerical codes. This section gives a brief description of these codes and the methods they adopt to solve the equations of ideal hydrodynamics. 

\subsection{Initial conditions} \label{ssec:ic_detail}
As a model for the hot cluster gas, we adopt the model of \cite{Komatsu2001} that we already described in \autoref{ssec:KS_model}. In our numerical experiments, we however make two important modifications compared to this model: First, we include the self-gravity of the gas sphere, which was ignored in the \citetalias{Komatsu2001} model. To this end, we rescale the gravitational acceleration of the model to that of only the dark matter by multiplying with a factor of $(1-M\sub{gas}/M\sub{100})$ and calculate the self-gravity of the gas self-consistently. Second, we adopt a polytropic exponent of $\gamma=5/3$, while \citetalias{Komatsu2001} originally adopted the effective polytropic index that they derived as a constraint of the model ($\gamma=1.137$ for our cluster parameters).

As a model for the bubble inflated by a central AGN, we use a sphere of radius $10\,{\rm kpc}$ placed close to the centre of the halo and heated to a temperature of $10^9\,{\rm K}$. The associated thermal energy, $\sim 3 \times 10^{59}$\,erg, is identical in all experiments and roughly consistent with what observations suggest \citep[e.g.][]{Birzan2004}.

\subsection{Lagrangian methods} \label{ssec:gizmo}
For all Lagrangian hydrodynamic simulations, we use the \gizmo\ code\footnote{publicly available at \url{https://bitbucket.org/phopkins/gizmo/wiki/Home}} \citep{Hopkins2015}, which includes various Lagrangian methods, among them TSPH, PSPH and MFM, that we use in what follows.

\begin{itemize}
\item Traditional SPH \citep[TSPH; e.g.][]{Lucy1977, Gingold1977, Monaghan1992} has been widely used in astrophysics, especially to study structure formation in the universe \citep[for recent reviews e.g.][and references therein]{Rosswog2009, Springel2010_SPH, Monaghan2012, Price2012} because of its great advantages, e.g. Galilean invariance, automatically adopted spatial resolution and exact mass conservation. However, it is also known that TSPH has difficulties to deal with fluid mixing. For example, the artificial tension on the contact surface of multi-phase fluids suppresses the growth of the KHI \citep[e.g.][]{Okamoto2003,Agertz2007}. 
\item Subsequent studies have made a lot of efforts to overcome the difficulties \citep[e.g.][]{Ritchie2001,Inutsuka2002,Read2010,Abel2011}. One of the modern formulations of SPH, the pressure flavour of SPH \citep[PSPH;][]{Saitoh2013, Hopkins2013} resolved them by replacing the volume element estimated from the mass density of an SPH particle (which is a technique used in TSPH) with that estimated from pressure (or energy density) of the particle and handled the fluid mixing instabilities, including the KHI and Rayleigh-Taylor instability. 
\item \cite{Hopkins2015} recently proposed a new class of particle methods for numerical hydrodynamics, meshless finite mass (MFM) and meshless finite volume (MFV), which have advantages of both SPH and grid-based schemes. These methods adopt a kernel-weighted volume discretization like SPH, but with a high-order matrix gradient estimator. A Riemann solver evaluates fluid (mass, momentum and energy) fluxes between particles, whose effective volume elements are overlapped. The limit of the MFM/MFV method with an infinitely sharply peaked kernel function corresponds to the moving-mesh method with non-regular deformed grids, e.g. the Voronoi tessellation \citep{Springel2010_Arepo,Duffell2011,Gaburov2012}.
\end{itemize}

The initial particle distribution which follows the \citetalias{Komatsu2001} density profile is drawn by using the rejection sampling scheme and thermal energy is assigned to each particle by interpolating the temperature profile of the \citetalias{Komatsu2001} model. To model AGN bubbles in the ICM, we increase the thermal energy of particles contained in the bubble within a radius of 10\,kpc to have temperature of $10^9$\,K. In all runs using \gizmo, we set the smoothing length, $h$, to the equivalent of what contains 32 neighbour particles and use the cubic spline kernel function. The gravitational softening in computing the gas self-gravity is fixed to be 1 kpc. We employ 67 108 864 particles, unless stated otherwise. The maximum spatial resolution is typically $2h \sim$ 10\,kpc at the centre and the mass resolution is $6.7 \times 10^{5}$\,\msun{}. The self-gravity of the gas is computed using the tree algorithm \citep{Barnes1986} with an opening angle of $\theta=0.7$ (default setting in \gizmo{}). The gravity of the DM halo is computed with a fixed analytical potential (see \autoref{ssec:KS_model} and \autoref{ssec:ic_detail} for details).

\subsection{Eulerian methods} \label{ssec:ramses}

For the Eulerian hydrodynamics simulations, we use the adaptive mesh-refinement (AMR) code \ramses{}\footnote{publicly available at \url{https://bitbucket.org/rteyssie/ramses}}. 
\ramses{} solves the hydrodynamic equations using a second-order, unsplit Godunov scheme. This method is known to accurately capture shocks. Fluxes are reconstructed from the cell-centred values with the Harten-Lax-van Leer-Contact (HLLC) Riemann solver that uses a first-order MinMod Total Variation Diminishing scheme. \ramses{} uses a tree structure, which allows for cell-by-cell refinement, thanks to which computational resources can be focused at high-density regions.

The initial conditions for the density and pressure of the ICM sphere are interpolated from the respective \citetalias{Komatsu2001} profiles. The AGN bubble is modelled by raising the temperature of all cells whose centres fall within the bubble radius to $10^9$\,K. We have used boundary conditions which allow only for outflow. Cells are refined based on the quasi-Lagrangian approach, when gas mass in the cell exceeds $1.94\times10^7$\,\msun{}. This leads to a similar number of leaf cells, 59 211 888 at the initial time, compared to the number of particles in GIZMO runs. Maximum spatial resolution achieved is 1.64\,kpc (11 levels of refinement), while minimum resolution is 52.5\,kpc (6 levels of refinement; base grid). We use analytical gravity for the DM halo described in \autoref{ssec:KS_model} and \autoref{ssec:ic_detail}. Self-gravity of the gas is calculated using the relaxation solver in \ramses{} and added to the halo potential.

\subsection{Spherical 1D validation code} \label{ssec:1d_code}

For spherically symmetric initial conditions in general cases where analytic solutions do not exist, we have run a simple spherical 1D MUSCL solver which includes solvers for one-dimensional spherical hydrodynamics and self-gravity. We use this 1D code in order to validate the solutions of the three-dimensional solvers discussed above during those stages when the solution is still close to spherically symmetric. The initial gas density and temperature are set as in \ramses. The boundary conditions are reflective in the centre and outflow at the outer boundary. Self gravity can be calculated trivially in spherical 1D by summing mass shells up to a given radius and the same analytical potential with that in runs of \gizmo{} and \ramses{} is adopted to compute the gravity of the DM halo. A more detailed description and tests for the 1D code can be found in \autoref{app:1d_solver}.

\section{Simulations} \label{sec:sims}
In what follows, we present the results of our numerical experiments. We first verify that our ICM is indeed close to hydrostatic equilibrium and remain so over an extended period of time with all numerical methods. Next, we investigate the evolution of an AGN-inflated hot bubble in such a hydrostatic ICM. We also consider the sophistication of our ICM model by replacing a fraction of its thermal energy with turbulence.

\subsection{Stability of the ambient gas sphere} \label{ssec:stab_test}
\begin{figure}
\begin{center}
\includegraphics[width=1\columnwidth]{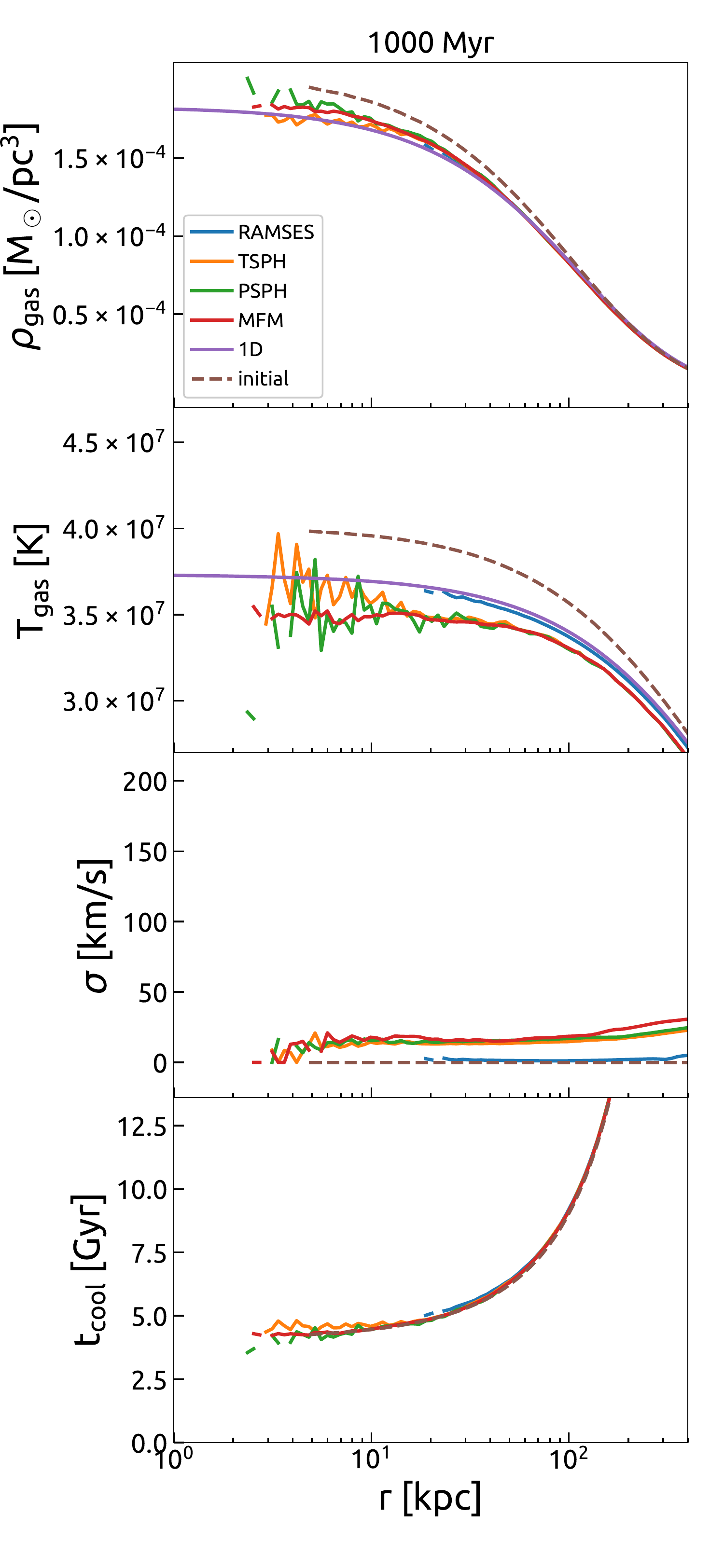}
\end{center}
\caption{
Radial profiles of gas density (first panel), temperature (second), velocity dispersion (third) and cooling timescale (fourth) after 1~Gyr of isolated evolution. Solid lines show profiles from numerical simulations using \ramses (blue), TPSH (yellow), PSPH (green), MFM (red) and the 1D code (purple), and the brown dashed one represents the profiles of the \citetalias{Komatsu2001} model. Each simulation is initialised in the same way - density and temperature follow \citetalias{Komatsu2001}. All runs are performed at a lower resolution than in the runs with an AGN bubble - GIZMO simulations with 8 388 608 particles, RAMSES simulation with \texttt{levelmax}=9. 
\label{fig:nofeedback_1000}}
\end{figure}

We first verify how close to hydrostatic equilibrium the clusters remain over an evolution time of 1~Gyr. The results of this stability test are shown as solid lines in \autoref{fig:nofeedback_1000}. Despite the differences from the original model of \citetalias{Komatsu2001} (self-gravity of the gas and equation of state), the differences between the original model (brown dashed line) and the simulation results are small. The gas profiles for density (first row) and temperature (second row) show a very minor evolution due to the system readjusting to a new hydrostatic equilibrium with an expansion of $\sim 30{\rm km/s}$. More remarkably, a similar degree of stability is seen when we replace some thermal energy with turbulent kinetic energy. And, even if we include turbulence at a very high level of $T\sub{nt} = 0.5T\sub{vir}$ in the ambient gas sphere, the spherically averaged radial profiles of density and temperature do not significantly deviate from the state shown in \autoref{fig:nofeedback_1000}. Here $T\sub{vir}$ is the virial temperature of the DM halo, $T\sub{vir}\approx 1.8 \times 10^7$\,K and the corresponding velocity dispersion is $\sigma\sub{vir}\sim835$\,km/s. We can therefore conclude that the gas sphere is reasonably stable and we adopt it as the ambient ICM of the AGN bubble in the following simulations. We note that our decision to neglect radiative cooling throughout this paper is well justified since the cooling timescale is longer than the buoyant timescale ($\sim1\,$Gyr), that we study here, by a factor of $\sim 5$, as shown in the fourth row.

In all cases we investigated, the turbulent energy decayed in a much shorter time, $\sim 30 {\rm Myr}$, than the total time of evolution of 1~Gyr. We also note that the Poisson particle noise puts the gas locally out of hydrostatic equilibrium in all runs using \gizmo\ . This drives a persistent `particle jitter' since the system responds by producing velocity dispersion (third row) which carries the difference in internal energy when compared to \ramses\ and the 1D code.

\subsection{Expansion of a central AGN bubble} 
\label{ssec:central_bubble}

\begin{figure}
\begin{center}
\includegraphics[width=1\columnwidth]{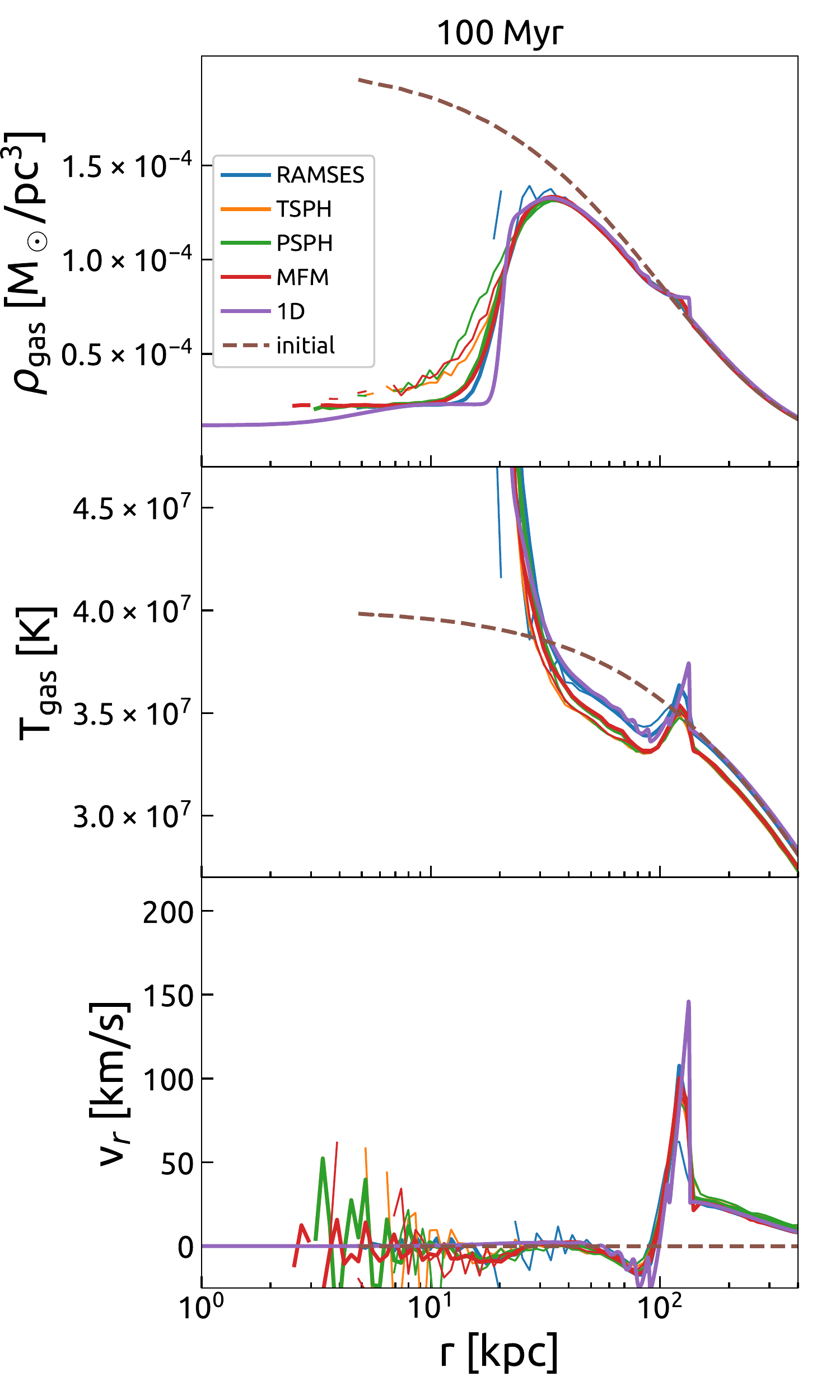}
\end{center}
\caption{
Radial profiles of gas density (upper), temperature (middle) and radial velocity (bottom)  100\,Myr after an injection of thermal energy by raising the temperature in the central 10\,kpc to $10^9$\,K. Solid lines show profiles from numerical simulations using \ramses (blue), TSPH (yellow), PSPH (green), MFM (red) and the 1D code (purple). Thick and thin lines are the results of simulations with the high- (64 million particles in the \gizmo\, runs and \texttt{levelmax}=11 in the \ramses\, run) and low resolution (8 million particles in the \gizmo\, runs and \texttt{levelmax}=9 in the \ramses\, run), respectively. The initial condition is shown as a brown dashed line. 
\label{fig:bub_cen_100_profiles}
}
\end{figure}

In this subsection, we study the expansion of an AGN bubble inflated at the centre of the ICM sphere. We increase the initial temperature of fluid elements, i.e. particles in the \gizmo\, runs and cells in \ramses{} and the 1D code, within a central spherical region of 10\,kpc to $10^9$\,K while the density profile follows that of the \citetalias{Komatsu2001} model.
The initial gas velocity is set to zero, i.e. we do not include turbulent velocities yet. While the central bubble is buoyantly unstable according to the Schwarzschild stability condition for convection, \autoref{eq:stab_cond}, the instability should not arise due to the symmetry of the system. In perfect symmetry it should just expand, but particle noise (\gizmo{}) and anisotropy due to the Cartesian mesh (\ramses{}) break this symmetry and let the bubble rise after some time (see \appref{app:images_bub_cen}). During the early phase of the simulations the bubble remains however symmetric and we can compare the results of the 3D simulations with that of 1D simulations with a much higher resolution. 

\autoref{fig:bub_cen_100_profiles} depicts radial profiles of gas density (upper), temperature (middle) and radial velocity (bottom) after 100\,Myr. The expanding bubble creates a strong shock wave which propagates outward. The shock positions in the simulations with different hydrodynamic solvers agree very well with each other ($\sim 100$\,kpc). The shock wave leaves a diffuse and hot core at the centre by compressing and accumulating the ambient ICM. While the overall features of the 1D solution are captured by all 3D runs, some differences in the core ($r\la10$\,kpc) are visible due to the lack of resolution in the 3D runs. The profiles obtained in 3D simulations are consistent with each other in the radial range of $r=10^1-10^3$\,kpc. 
Comparing the results of 3D simulations at high (thick lines) and low (thin lines) resolution, the profiles are numerically converged down to $r \sim 20$\,kpc, which corresponds to the spatial resolution in the simulations with lower resolution. An additional interesting difference between the 1D and 3D runs is the presence of a pulsating mode in the 1D run interior to the shock and visible as wiggles behind the shock position. It appears that due to lack of resolution such modes are efficiently damped out in the 3D runs.

\subsection{A rising bubble in non-turbulent ICM}
\label{ssec:shifted_bubble_wo_turbulence}

In this subsection, we investigate the rising of buoyantly unstable AGN bubbles and their interaction with the ambient ICM. As in the simulations of the previous section, we change the initial setup by increasing the temperature of fluid elements within the bubble with a radius of 10\,kpc to $10^9$\,K. Now however, the centre of the bubble is shifted to the upper right oblique 45 degree direction in the $x-y$ plane by 10\,kpc from the centre, keeping the temperature outside the bubble and density to be those of the \citetalias{Komatsu2001} model. Turbulence is not taken into account and the initial velocity is set to be zero. Since the shifted hot bubble breaks both the Schwarzschild stability condition for convection, \autoref{eq:stab_cond}, and the symmetry of the system, it must be buoyantly unstable and rising. The amount of the injected thermal energy is almost the same as that in the simulations with a {\em central} bubble, $\sim 3 \times 10^{59}$\,erg. 

\begin{figure*}
\begin{center}
\includegraphics[width=0.95\textwidth]{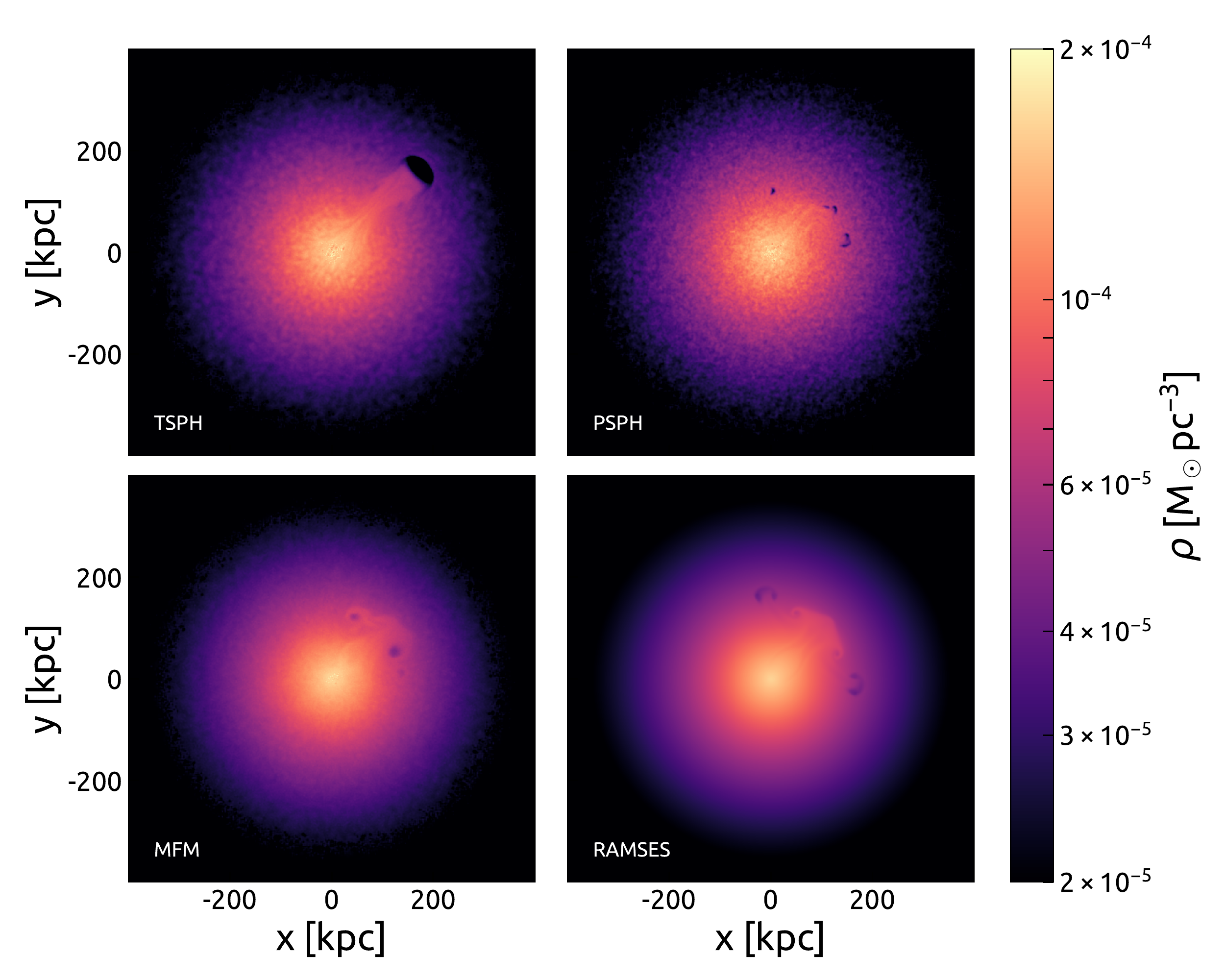}
\end{center}
\caption{
Slices through the gas density distribution for TSPH (top left), PSPH (top right), MFM (bottom left) and RAMSES (bottom right) after 1\,Gyr of evolution of a bubble initially shifted from the centre. Each simulation is initialised in the same way - density and temperature follow \citetalias{Komatsu2001} and we artificially raise the temperature within the bubble whose centre is shifted to the upper right oblique 45 degree direction in the $x-y$ plane by 10\,kpc from the centre of the gas sphere to $10^9$\,K. The radius of the bubble, 10\,kpc, is the same as the runs shown in the previous section. 
\label{fig:bub_shift_1000}}
\end{figure*}

\autoref{fig:bub_shift_1000} shows slices of gas density after 1\,Gyr in the runs with the shifted bubble. The fate of rising bubbles very clearly depends strongly on the choice of hydrodynamical solvers. The bubble is rising towards the direction of the initial displacement (upper right oblique 45 degree direction in the $x-y$ plane). It survives unharmed and reaches large radii in the TSPH run (upper left). This result is inconsistent with the analytical expectation discussed in \autoref{fig:tbuoyancy_vs_tkhi} and caused by the well-known suppression of fluid instabilities by the spurious surface tension \citep[e.g.][]{Okamoto2003,Agertz2007}. In other runs, the bubble is dissolved by the KHI while configurations are different from one another. The bubble is fragmented into smaller ones in the PSPH run (upper right). In contrast, the MFM one (lower left) looks very similar to the \ramses\ result (lower right), while the symmetry of the structure is broken in the MFM run by Poisson noise contained in the initial particle distribution (see also \appref{app:resoution} for a discussion of resolution effects). The difference in bubble morphology between hydrodynamical solvers of  translates of course directly into differences in the redistribution of mass, including metals, and energy by the rising bubble. 

\begin{figure}
\begin{center}
\includegraphics[width=0.5\textwidth]{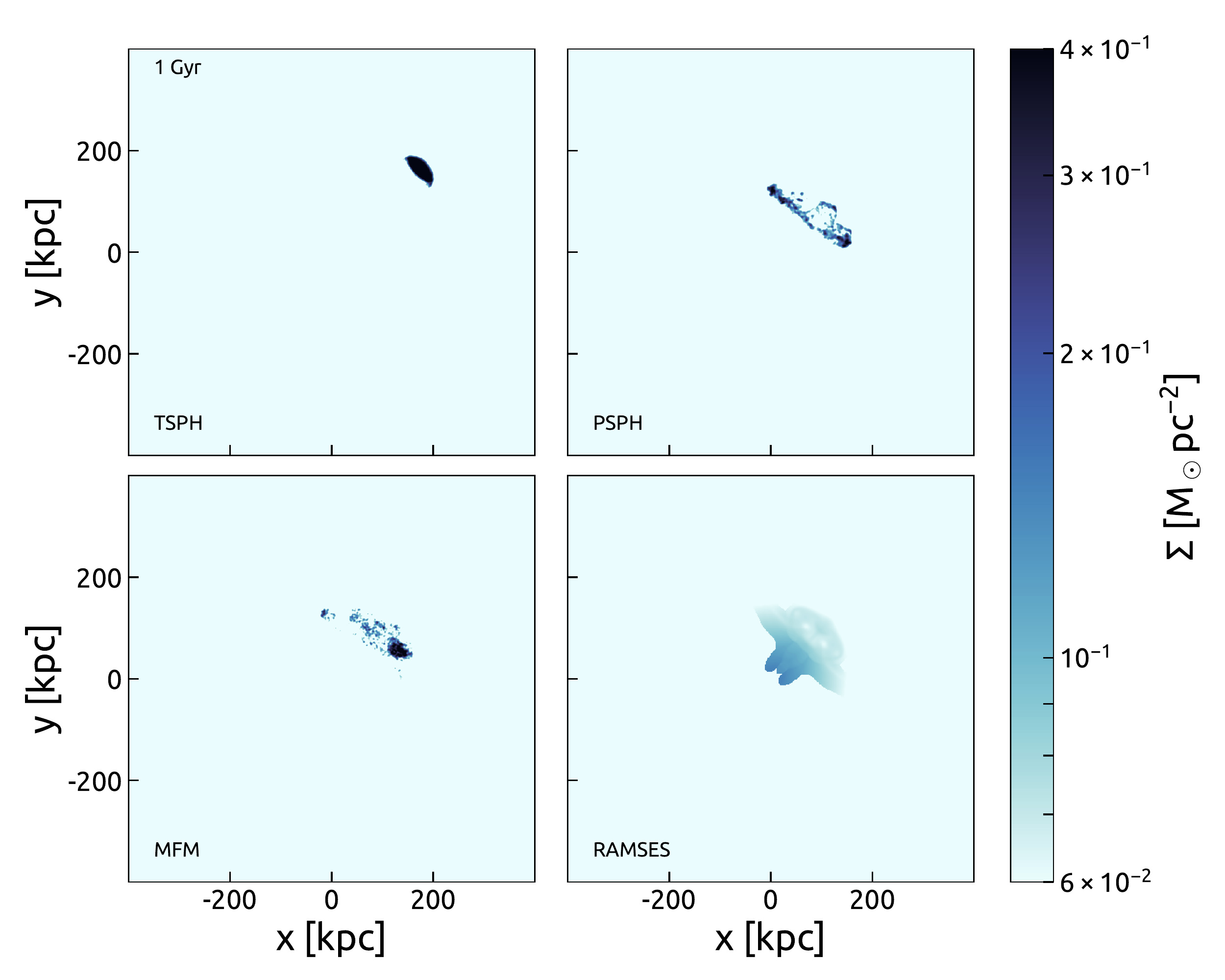}
\end{center}
\caption{
Projection of the density distribution of gas initially contained in the heated bubble for TSPH (top left), PSPH (top right), MFM (bottom left) and RAMSES (bottom right) after 1\,Gyr of evolution. The same simulations as in \autoref{fig:bub_shift_1000} are shown. 
\label{fig:bub_mix_nt}}
\end{figure}

In order to better visualize the fate of the heated fluid elements, we next study how the rising bubble is dissolved in the ICM in more detail by explicitly tracking the fluid elements initially contained in the heated sphere. For the \gizmo{} runs (TSPH, PSPH and MFM), the Lagrangian nature allows us to follow the heated fluid elements through the ID of particles. If metal mixing is not explicitly performed in SPH, this would also track the evolution of metals contained in the hot bubble. While MFM method computes the fluxes between particles like Eulerian schemes and  thus gas initially contained in the bubble may dissipate, we use the ID of particles to follow the heated fluid elements. In Eulerian schemes, like \ramses{}, tracking fluid elements is more complicated. We take a simple approach here and follow the gas that was initially within the bubble by injecting a passive tracer variable in the grid, which is advected with the flow of the gas. This is similar in spirit to injecting metals from supernovae explosions, but in our case we inject the tracer only at the beginning of the simulation and its value in no way modifies any other properties of the grid. This method only approximates tracking of the gas flow in the grid and a more advanced method which uses tracer particles in \ramses\ is currently in development (Cadiou et al., private communication). Having that in mind we caution that direct comparison between grid passive tracer and particle IDs can be only qualitative. On the other hand, both cases in fact correspond rather well to the evolution of metals in the flow in the different methods.

\autoref{fig:bub_mix_nt} shows the projected density, integrated along the line of sight (LoS), of gas initially contained in the heated bubble at 1\,Gyr. It is apparent that, in all cases, the bubble is not fully mixed with the ICM yet, while the specific bubble configuration depends on the choice of hydrodynamical solver, just like the density distribution of the ICM. In the TSPH run (upper left), the heated fluid elements are confined to a small region which corresponds to the less-dense cavity in the upper left panel in \autoref{fig:bub_shift_1000}. This is another indication that the bubble survives largely unaffected in the TSPH run due to the artificial suppression of the growth of fluid instabilities. In other runs, the bubble is more elongated and less-dense compared with the TSPH run since they handle the instabilities better while the bubble is not completely dissolved and mixed with the ICM even in these runs. In the PSPH (upper right) and MFM (lower left) runs, one can find small density fluctuations on the bubble surface which originate from the Poisson noise in drawing of the initial particle distribution. The bubble has a symmetric structure in the \ramses{} run thanks to the absence of such noise. Fluctuations in density and velocity fields would of course always exist in the real ICM. We will study the impact of such `noises' below in \autoref{ssec:shifted_bubble_w_turbulence} by introducing, in a controlled way, a turbulent velocity field. 

\begin{figure*}
\includegraphics[width=1\textwidth,height=0.8\textheight]{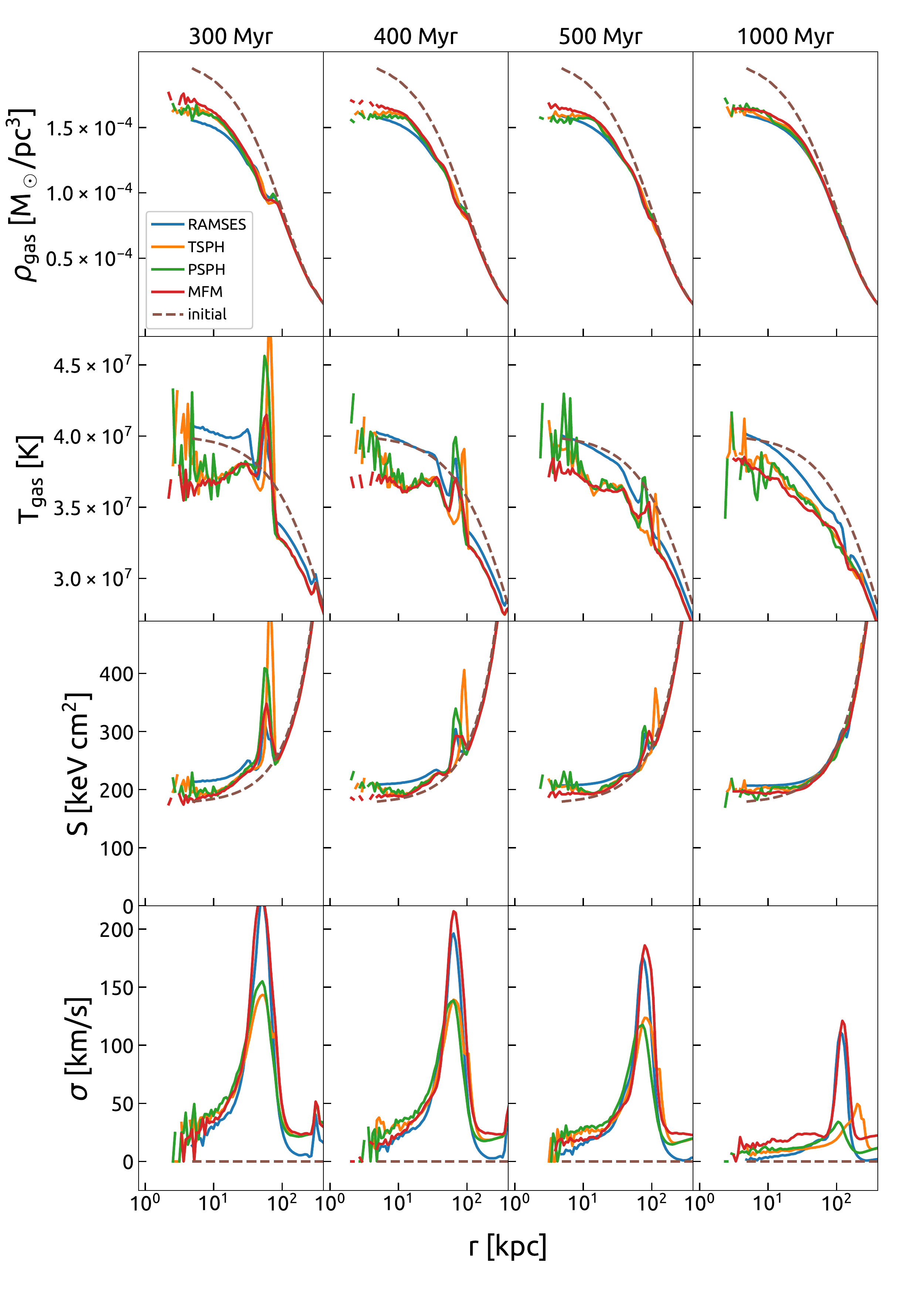}
\caption{
Mass-weighted profiles after evolution of 300\,Myr (first column), 400\,Myr (second), 500\,Myr (third) and 1\,Gyr (fourth). We focus our analysis only on the quarter of the box in which the bubble evolves. The profiles of gas density (first row), temperature (second), entropy (third) and three-dimensional velocity dispersion (fourth) are shown. Solid lines show profiles from numerical simulations using \ramses{} (blue), TPSH (yellow), PSPH (green) and MFM (red). The brown dashed line represents the \citetalias{Komatsu2001} model. 
\label{fig:bub_shift_profiles}}
\end{figure*}

The pressing question is of course whether such morphological differences are also reflected in integral properties of the ICM and thus affect the efficiency of AGN feedback. We thus next investigate how the bubble redistributes mass and energy into the ICM in \autoref{fig:bub_shift_profiles}. The evolution of the profiles obtained from the quarter of the box in which the bubble rises is drawn from the left-most to right-most columns. As shown in \autoref{fig:bub_cen_100_profiles}, the blast wave ignited by the injected thermal energy reaches to $r \ga 100$\,kpc at $t=100$\,Myr. In those runs the energy was damped at the centre of the gas sphere. Assuming the self-similarity of the Sedov-Taylor solution and that the blast wave expands analogously in the runs in which the bubble is shifted, then it would reach $r=200-300$\,kpc at $t=300$\,Myr. Indeed, as expected, it appears as peaks in the outskirts ($r \sim 300$\,kpc) in the profiles of temperature (second row) and velocity dispersion (fourth row) of the left-most column. At later times it passes through the radial range demonstrated in \autoref{fig:bub_shift_profiles}. Hence the dominating peaks shown in this figure are not originated by the blast wave, but by the rising bubble.

The bubble appears as the bump in the density profile (first row) and as peaks in others. Remarkably, we do not observe significant differences in the density profile. As illustrated in \autoref{fig:bub_shift_1000}, the hot bubble robustly survives for a long time in the TSPH run because of the spurious tension and we see rising of the hot bubble in the profiles of temperature (second row) and entropy (third row). On the other hand, because the bubble is being dissolved and turbulent motions could arise in MFM and \ramses{} runs, the thermal energy contained in the bubble is converted into kinetic energy (fourth row) and the peaks are less pronounced in the profiles of temperature and entropy. The behaviour of the PSPH run is intermediate between the two groups, as expected from \autoref{fig:bub_shift_1000} in which the bubble is fragmented into smaller ones in the PSPH run. Note the velocity dispersion, 100-200 km/s, of the turbulent motion driven by the bubble is roughly consistent with the observation of the Perseus cluster \citep{Hitomi2016} and simulation results by \citet[][but see also \citealt{Reynolds2015}]{Lau2017}.

In summary, the choice of hydrodynamical solver does change the fate of the buoyantly unstable hot bubble and the mass and energy redistribution driven by it in simulations. In the TSPH run, the rising bubble survives for a time inconsistent with the analytical expectation and reaches large radii because spurious tension suppresses the growth of fluid instabilities on the surface. As a result, the thermal energy is locked in the bubble and a smaller fraction of energy, compared with those in the runs using the other hydrodynamical solvers, is distributed to the ambient medium. On the other hand, in the simulations using hydrodynamical solvers which can handle fluid instabilities, the bubble is dissolved in a timescale consistent with the analytical expectation (a few 100\,Myr, see \autoref{fig:tbuoyancy_vs_tkhi}) and the thermal energy originally contained in the bubble is converted to non-thermal turbulent energy. In the real ICM, energy would continuously change the form, e.g. transformation from thermal energy to kinetic energy (via bubble rising) and vice versa (via dissipation). 

In addition, the different evolution of the bubble can also lead to differences in the metal distribution in the ICM. Supposing that the bubble, which comes from the centre of the cluster, is metal enriched (if metallicity gradients are present), a more diffusive metal distribution, i.e. lower metallicity, would be observed in simulations using the schemes which can handle fluid instabilities \citep[see e.g.][]{Martizzi2016}. This would be further enhanced if the metals are diffused between fluid elements. We study the more complicated and realistic phenomena, rising of an AGN bubble in a turbulent ICM, in the next section.

\subsection{A rising bubble in a turbulent ICM}
\label{ssec:shifted_bubble_w_turbulence}

\begin{figure}
\begin{center}
\includegraphics[width=0.5\textwidth]{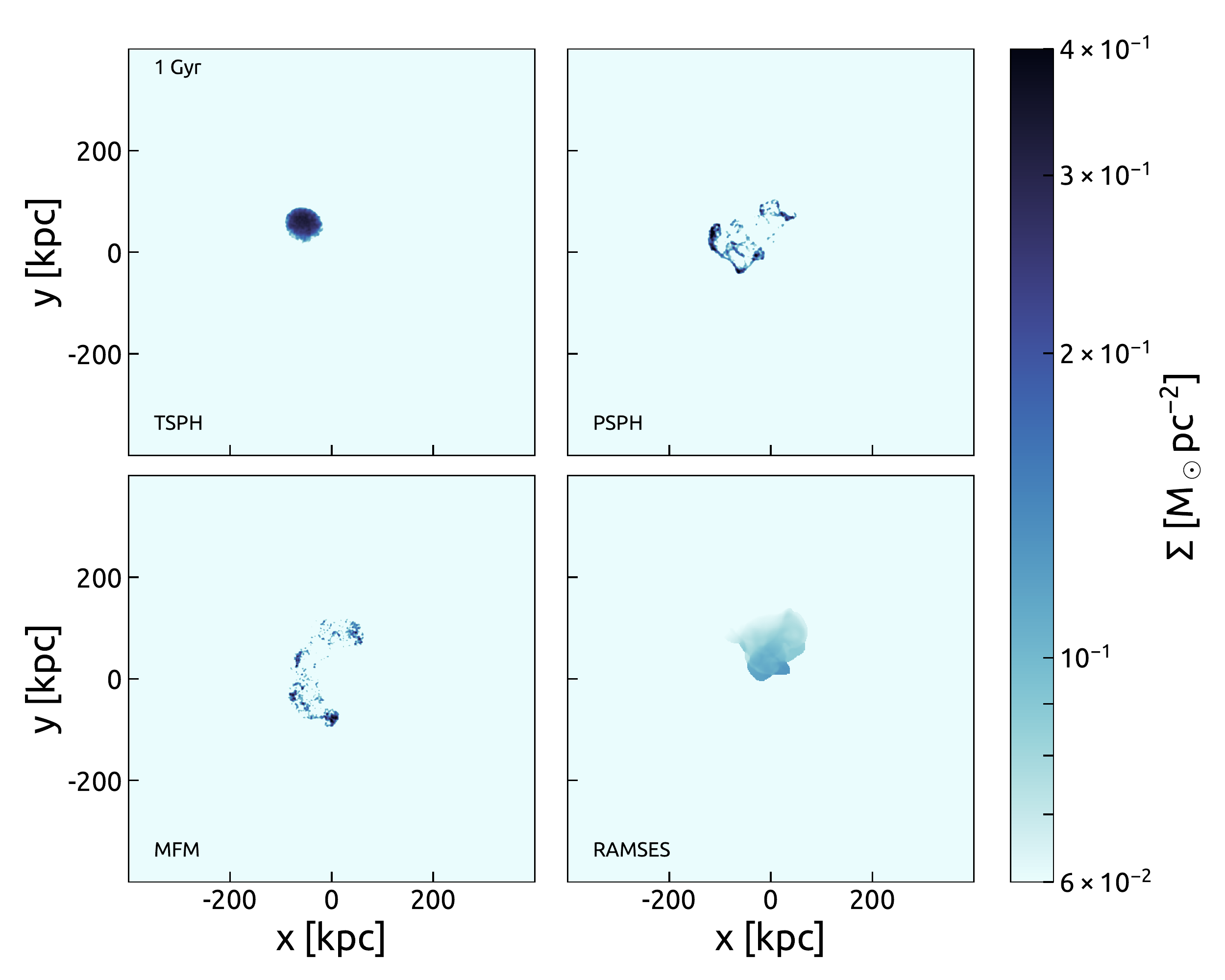}
\end{center}
\caption{
Projection of the density distribution of gas initially contained in the heated bubble for TSPH (top left), PSPH (top right), MFM (bottom left) and RAMSES (bottom right) after 1\,Gyr of evolution. Here, the results of simulations with a controlled turbulent velocity field of $T\sub{nt}=5.4\times10^5$\,K are shown. 
\label{fig:bub_mix_t}}
\end{figure}

This subsection investigates how small fluctuations in the ICM affect the rising bubble and also their back-reaction. In order to study this effect, we introduce a turbulent velocity field following a Kolmogorov power spectrum (see \autoref{ssec:int_turb}) with the non-thermal temperature of $T\sub{nt} = 0.03T\sub{vir} \approx 5.4 \times 10^5 {\rm K}$ which corresponds to a velocity dispersion of 147km/s, consistent with the X-ray observation of the Perseus cluster (\citealt{Hitomi2016}, see also \citealt{ZuHone2017}). In the initial conditions of the ICM we subtract the corresponding thermal energy from the ICM. Other parameters such as resolution parameters, bubble temperature and size, are the same as those in \autoref{ssec:shifted_bubble_wo_turbulence}. 

\autoref{fig:bub_mix_t} shows the projected density distribution of gas initially contained in the heated bubble at 1\,Gyr in the runs with the turbulent velocity field. Because of the perturbations due to turbulence, the bubble does not rise in the direction of the initial offset (upper right oblique 45 degree direction in the $x-y$ plane). In addition, turbulence introduces perturbations of smaller wavelengths which grow faster (see \autoref{eq:tau_khi}) and thus enhances instability of the bubble surface. As a result, the bubble is more quickly disrupted and mixed with the ICM compared to the non-turbulent case (\autoref{fig:bub_mix_nt}), using hydrodynamical schemes which can handle the instabilities (PSPH, MFM and \ramses{}). However, just as in the non-turbulent run, the bubble in the TSPH run robustly survives due to the suppression of the growth of fluid instabilities by artificial surface tension (upper left). Note that the distance from the centre of the cluster to the (remnants of) bubbles in the turbulent runs is smaller than those in the non-turbulent ones. This and enhanced mixing imply that redistribution of mass, including metals, and energy may be also enhanced around the centre in this case. 

\begin{figure*}
\includegraphics[width=1\textwidth,height=0.8\textheight]{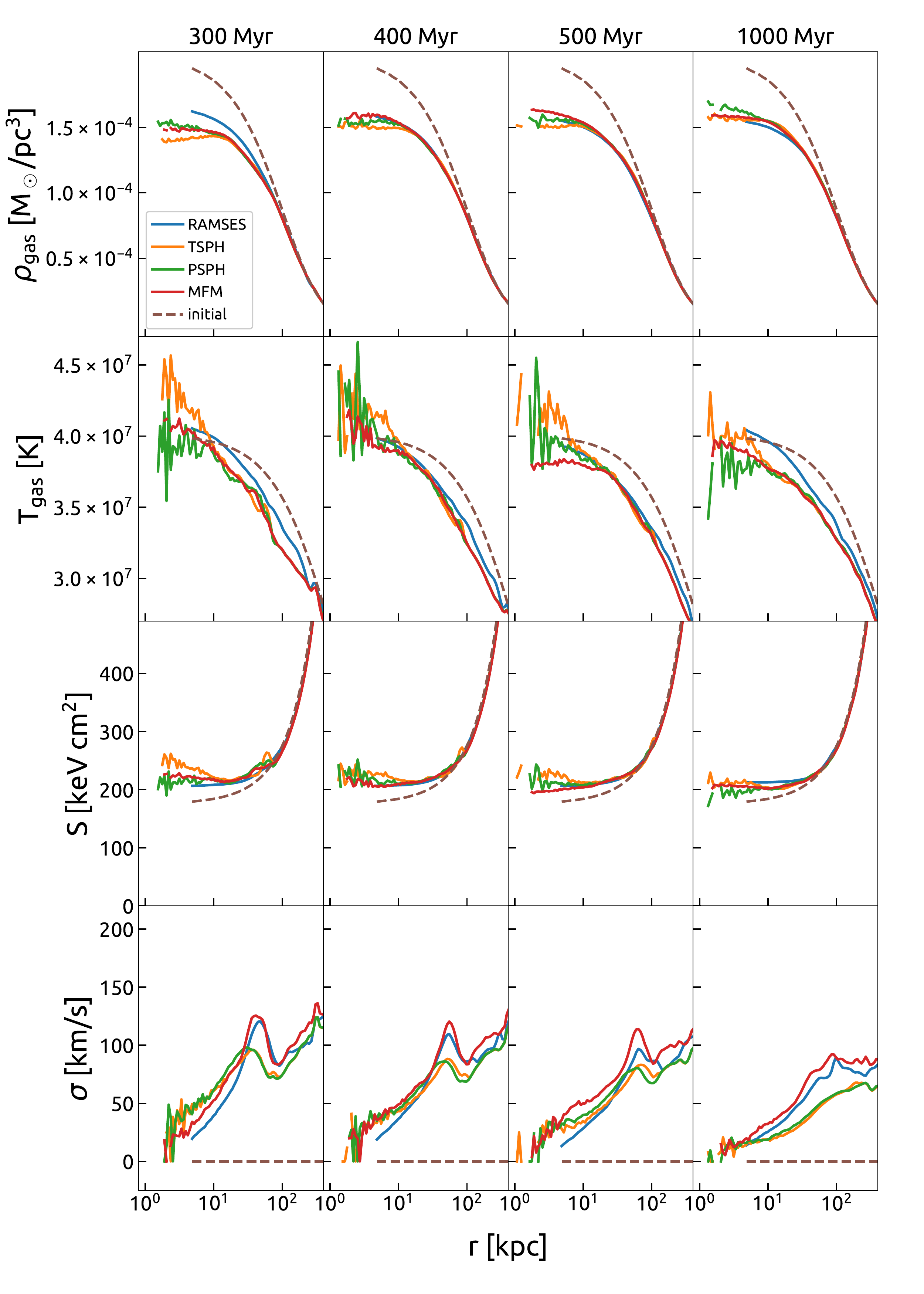}
\caption{
Mass-weighted profiles after 300\,Myr (first column), 400\,Myr (second), 500\,Myr (third) and 1\,Gyr (fourth) with turbulent initial conditions. Here, we analyse the whole box in each simulation. The profiles of gas density (first row), temperature (second), entropy (third) and three-dimensional velocity dispersion (fourth) are shown. Solid lines show profiles from numerical simulations using \ramses{} (blue), TPSH (yellow), PSPH (green) and MFM (red). The brown dashed line represents the \citetalias{Komatsu2001} model. 
\label{fig:turb_gas_profiles}}
\end{figure*}

We study this point using \autoref{fig:turb_gas_profiles} which presents the profiles of gas density (first row), temperature (second), entropy (third) and velocity dispersion (fourth). The evolution is shown from the left-most to right-most columns. Unlike before, we now use the whole simulation box to measure the profiles since the direction of rising of the bubble is no longer the same in all runs, as shown in \autoref{fig:bub_mix_t}. We note that this dilutes the signature of bubbles in the profiles since the part of the simulation box that is not significantly affected by the bubble is included in the analysis. We see no significant differences between hydrodynamical solvers in the first three profiles from the top. However, peaks in the velocity dispersion profiles are more evident in the MFM and \ramses{} runs compared with the SPH runs. Again, we interpret this as differences in the efficiency of conversion of thermal energy to turbulent (kinetic) energy when the bubble rises. 

It is also worth mentioning that the turbulence decays in all runs, especially in the centre of the cluster where the crossing time is shorter than that measured at the outskirts. Interestingly, the decay of the turbulent velocity field is more significant in the SPH runs compared with the MFM and \ramses{} ones. Since perturbations with shorter wavelength, which are introduced by turbulence and grow faster, enhance dissolving of the bubble and mixing with the ICM, these mechanisms would be non-linearly degraded in SPH simulations, while a stronger turbulent velocity field is kept in the MFM and \ramses{} runs. We expect that the sub-grid physics models in cosmological simulations which trigger sequential and/or multiple AGN bubbles in the ICM can enhance such contrasts between hydrodynamical solvers. To test this hypothesis, more systematic comparison studies varying not only hydrodynamical solvers but also resolution and sub-grid physics would be needed, which is however beyond the scope of this brief study.

\section{Summary and discussion}
\label{sec:summary_discussion}

AGN feedback is believed to be an important heating source to keep the ICM in the cluster centre hot and diffuse, and a key to avoid the central cooling flow problem. A number of studies have used numerical simulations to investigate the effect of AGN feedback on the properties of the ICM. However, despite progress in hydrodynamical solvers and modelling of AGN feedback in numerical simulations, a full consensus among cosmological hydrodynamical simulations has not been reached yet and simulations for cluster cosmology are not yet predictive. This situation motivated us to investigate one of the possible reasons for the inconsistency: the difference in the hydrodynamical schemes employed in various simulations. For this purpose, we studied the evolution of rising bubbles inflated by AGN feedback in ideal self-gravitating hydrodynamics. According to  observations, the bubbles may be pockets of a significant amount of thermal energy (and cosmic rays, which are however not yet routinely included in cosmological simulations) and hence studying the interaction between the bubble and the ICM is important to make progress. 

Using a simplified model, we first showed that a hot bubble will rise buoyantly in the ICM and is prone to surface instabilities that act to disrupt and mix it over time scales of 1~Gyr for a typical cluster of a few $10^{14}$\,\msun{}. If the hydrodynamical solver captures fluid instabilities (i.e. the KHI here) well (such as MFM and \ramses{} that we studied here), the bubble is disrupted as expected. However, the bubble survives for a longer time than the analytical expectation in simulations using traditional SPH due to spurious surface tension suppressing the growth of instabilities. In the simulation employing the PSPH scheme, we observed that the bubble instead fragments into smaller ones. In addition, we found that if the bubble is metal enriched compared to the ICM at larger radii, a more diffusive metal distribution, i.e. lower metallicity, may be observed in simulated galaxy clusters using hydrodynamical solvers which can model well fluid instabilities \citep[see also e.g.][]{Martizzi2016}, and particularly so in when comparing with Lagrangian schemes that in addition neglect metal diffusion at the fluid element scale.

Because of the difference in bubble mixing, the energy redistribution in the ICM driven by the rising bubble depends somewhat, but less than initially expected, on the choice of hydrodynamical solvers in simulations. Using the hydrodynamical solvers that can capture fluid instabilities, the thermal energy contained in the bubble is transformed more readily into kinetic energy because dissolving of the bubble drives turbulent motion in the ICM. In contrast, the thermal energy is captured in the surviving bubble and transported to large radii in the TSPH simulation. The energy redistribution observed in the PSPH simulation is similar to that in the TSPH run.

In a second step, we investigated whether the various methods agree better in a turbulent ICM, where random motion act to increase bubble disruption and mixing. When the numerical method resolves well fluid instabilities (here MFM and \ramses{}), the thermal energy initially contained in the hot bubble is again efficiently converted to kinetic energy of the turbulent motion. More surprisingly, even if we include the turbulent velocity field, the bubble robustly survived in the TSPH simulation while it is more significantly dissolved and mixed with the ICM in all other hydrodynamical solvers. We also found that in general the SPH schemes used in this study damp the turbulent velocity field more strongly than both \ramses{} and MFM. And, they also lead to less efficient conversion of the hot bubble to turbulent motion. 

In summary, we have observed significant differences in the spatial distribution of the hot bubble after it mixes with the ICM when employing different hydrodynamical solvers. Somewhat surprisingly however, the differences in the gas profiles are not very significant. While these results might imply that the choice of the hydrodynamical solvers is not the primary factor to explain differences in results obtained in different cosmological hydrodynamical simulations, we can only speculate here whether the differences we observed may be amplified if AGN feedback energy is injected repeatedly as the halo grows in mass. The amount of injected energy into the AGN bubble in our simulations, $\sim 10^{59}$\,erg, was only $\sim 0.1\%$ of the total thermal energy of the ICM. Assuming that the central super massive black hole, which is the engine of the AGN, has a mass of $10^9$ \msun{}, the feedback energy over the timescale of our simulations, 1\,Gyr, can be of order of $\sim 10^{62-63}$\,erg if the AGN continues to be active \citep[see e.g.][]{Woo2002}. While this na\"ive estimation provides of course just an upper limit, the budget of  feedback energy could be much greater than what we assumed and recursive injections of AGN bubbles are expected in a duty cycle behaviour of central fuel gas consumption and replenishment. 

Even if the feedback energy is the same as assumed in this work, the impact of the bubbles may be greater if the ICM has a low-temperature core in the centre, i.e. in a cool-core cluster, so that the central gas is close to a runaway cooling instability.
We will revisit this aspect in future work.

Last but not least, additional physical processes that this study does not take into account, such as magnetic fields and cosmic rays, may alter the evolution of the AGN bubble and the thermodynamics galaxy clusters. In particular, one would expect that magnetic tension acts to suppress the disruption and mixing of buoyantly rising bubbles. We will investigate the interactions between the AGN bubble and ICM using idealised MHD simulations of several kinds of hydrodynamical solvers in the subsequent paper (Biernacki et al., in prep.).

\section*{Acknowledgements}
We would like to thank 
Arif Babul, 
Alessandro Lupi, 
Ewald Puchwein,
and 
Joakim Rosdahl 
for stimulating conversations. 
We also thank Daisuke Nagai for his careful reading of the draft and providing insightful comments. 
This work heavily relied on the \gizmo{} \citep{Hopkins2015} and \ramses{} \citep{Teyssier2002} codes and \textsc{pynbody} package \citep{pynbody}. We would like to thank the authors for their efforts. 
GO and OH acknowledge funding from the European Research Council (ERC) under the European Union's Horizon 2020 research and innovation programme (grant agreement No. 679145, project `COSMO-SIMS').




\bibliographystyle{mnras}
\bibliography{bibliography1} 

\begin{thebibliography}{}
\makeatletter
\relax
\def\mn@urlcharsother{\let\do\@makeother \do\$\do\&\do\#\do\^\do\_\do\%\do\~}
\def\mn@doi{\begingroup\mn@urlcharsother \@ifnextchar [ {\mn@doi@}
  {\mn@doi@[]}}
\def\mn@doi@[#1]#2{\def\@tempa{#1}\ifx\@tempa\@empty \href
  {http://dx.doi.org/#2} {doi:#2}\else \href {http://dx.doi.org/#2} {#1}\fi
  \endgroup}
\def\mn@eprint#1#2{\mn@eprint@#1:#2::\@nil}
\def\mn@eprint@arXiv#1{\href {http://arxiv.org/abs/#1} {{\tt arXiv:#1}}}
\def\mn@eprint@dblp#1{\href {http://dblp.uni-trier.de/rec/bibtex/#1.xml}
  {dblp:#1}}
\def\mn@eprint@#1:#2:#3:#4\@nil{\def\@tempa {#1}\def\@tempb {#2}\def\@tempc
  {#3}\ifx \@tempc \@empty \let \@tempc \@tempb \let \@tempb \@tempa \fi \ifx
  \@tempb \@empty \def\@tempb {arXiv}\fi \@ifundefined
  {mn@eprint@\@tempb}{\@tempb:\@tempc}{\expandafter \expandafter \csname
  mn@eprint@\@tempb\endcsname \expandafter{\@tempc}}}

\bibitem[\protect\citeauthoryear{{Abel}}{{Abel}}{2011}]{Abel2011}
{Abel} T.,  2011, \mn@doi [\mnras] {10.1111/j.1365-2966.2010.18133.x}, \href
  {http://adsabs.harvard.edu/abs/2011MNRAS.413..271A} {413, 271}

\bibitem[\protect\citeauthoryear{{Agertz} et~al.,}{{Agertz}
  et~al.}{2007}]{Agertz2007}
{Agertz} O.,  et~al., 2007, \mn@doi [\mnras]
  {10.1111/j.1365-2966.2007.12183.x}, \href
  {http://cdsads.u-strasbg.fr/abs/2007MNRAS.380..963A} {380, 963}

\bibitem[\protect\citeauthoryear{{Barnes} \& {Hut}}{{Barnes} \&
  {Hut}}{1986}]{Barnes1986}
{Barnes} J.,  {Hut} P.,  1986, \mn@doi [\nat] {10.1038/324446a0}, \href
  {http://cdsads.u-strasbg.fr/abs/1986Natur.324..446B} {324, 446}

\bibitem[\protect\citeauthoryear{{Barnes} et~al.,}{{Barnes}
  et~al.}{2017a}]{Barnes2017_IllustrisTNG}
{Barnes} D.~J.,  et~al., 2017a, preprint, \href
  {http://cdsads.u-strasbg.fr/abs/2017arXiv171008420B} {} (\mn@eprint {arXiv}
  {1710.08420})

\bibitem[\protect\citeauthoryear{{Barnes} et~al.,}{{Barnes}
  et~al.}{2017b}]{Barnes2017_ClusterEAGLE}
{Barnes} D.~J.,  et~al., 2017b, \mn@doi [\mnras] {10.1093/mnras/stx1647}, \href
  {http://cdsads.u-strasbg.fr/abs/2017MNRAS.471.1088B} {471, 1088}

\bibitem[\protect\citeauthoryear{{Battaglia}, {Bond}, {Pfrommer}  \&
  {Sievers}}{{Battaglia} et~al.}{2013}]{Battaglia2013}
{Battaglia} N.,  {Bond} J.~R.,  {Pfrommer} C.,   {Sievers} J.~L.,  2013,
  \mn@doi [\apj] {10.1088/0004-637X/777/2/123}, \href
  {http://cdsads.u-strasbg.fr/abs/2013ApJ...777..123B} {777, 123}

\bibitem[\protect\citeauthoryear{{B{\^i}rzan}, {Rafferty}, {McNamara}, {Wise}
  \& {Nulsen}}{{B{\^i}rzan} et~al.}{2004}]{Birzan2004}
{B{\^i}rzan} L.,  {Rafferty} D.~A.,  {McNamara} B.~R.,  {Wise} M.~W.,
  {Nulsen} P.~E.~J.,  2004, \mn@doi [\apj] {10.1086/383519}, \href
  {http://cdsads.u-strasbg.fr/abs/2004ApJ...607..800B} {607, 800}

\bibitem[\protect\citeauthoryear{{Booth} \& {Schaye}}{{Booth} \&
  {Schaye}}{2009}]{Booth2009}
{Booth} C.~M.,  {Schaye} J.,  2009, \mn@doi [\mnras]
  {10.1111/j.1365-2966.2009.15043.x}, \href
  {http://cdsads.u-strasbg.fr/abs/2009MNRAS.398...53B} {398, 53}

\bibitem[\protect\citeauthoryear{{Br{\"u}ggen} \& {Kaiser}}{{Br{\"u}ggen} \&
  {Kaiser}}{2002}]{Bruggen2002}
{Br{\"u}ggen} M.,  {Kaiser} C.~R.,  2002, \mn@doi [\nat] {10.1038/nature00857},
  \href {http://cdsads.u-strasbg.fr/abs/2002Natur.418..301B} {418, 301}

\bibitem[\protect\citeauthoryear{{Cavagnolo}, {McNamara}, {Nulsen}, {Carilli},
  {Jones}  \& {B{\^i}rzan}}{{Cavagnolo} et~al.}{2010}]{Cavagnolo2010}
{Cavagnolo} K.~W.,  {McNamara} B.~R.,  {Nulsen} P.~E.~J.,  {Carilli} C.~L.,
  {Jones} C.,   {B{\^i}rzan} L.,  2010, \mn@doi [\apj]
  {10.1088/0004-637X/720/2/1066}, \href
  {http://cdsads.u-strasbg.fr/abs/2010ApJ...720.1066C} {720, 1066}

\bibitem[\protect\citeauthoryear{{Churazov}, {Br{\"u}ggen}, {Kaiser},
  {B{\"o}hringer}  \& {Forman}}{{Churazov} et~al.}{2001}]{Churazov2001}
{Churazov} E.,  {Br{\"u}ggen} M.,  {Kaiser} C.~R.,  {B{\"o}hringer} H.,
  {Forman} W.,  2001, \mn@doi [\apj] {10.1086/321357}, \href
  {http://cdsads.u-strasbg.fr/abs/2001ApJ...554..261C} {554, 261}

\bibitem[\protect\citeauthoryear{{Cowie} \& {Binney}}{{Cowie} \&
  {Binney}}{1977}]{Cowie1977}
{Cowie} L.~L.,  {Binney} J.,  1977, \mn@doi [\apj] {10.1086/155406}, \href
  {http://cdsads.u-strasbg.fr/abs/1977ApJ...215..723C} {215, 723}

\bibitem[\protect\citeauthoryear{{Dennis} \& {Chandran}}{{Dennis} \&
  {Chandran}}{2005}]{Dennis2005}
{Dennis} T.~J.,  {Chandran} B.~D.~G.,  2005, \mn@doi [\apj] {10.1086/427424},
  \href {http://cdsads.u-strasbg.fr/abs/2005ApJ...622..205D} {622, 205}

\bibitem[\protect\citeauthoryear{{Di Matteo}, {Springel}  \& {Hernquist}}{{Di
  Matteo} et~al.}{2005}]{DiMatteo2005}
{Di Matteo} T.,  {Springel} V.,   {Hernquist} L.,  2005, \mn@doi [\nat]
  {10.1038/nature03335}, \href
  {http://cdsads.u-strasbg.fr/abs/2005Natur.433..604D} {433, 604}

\bibitem[\protect\citeauthoryear{{Di Matteo}, {Khandai}, {DeGraf}, {Feng},
  {Croft}, {Lopez}  \& {Springel}}{{Di Matteo} et~al.}{2012}]{DiMatteo2012}
{Di Matteo} T.,  {Khandai} N.,  {DeGraf} C.,  {Feng} Y.,  {Croft} R.~A.~C.,
  {Lopez} J.,   {Springel} V.,  2012, \mn@doi [\apjl]
  {10.1088/2041-8205/745/2/L29}, \href
  {http://cdsads.u-strasbg.fr/abs/2012ApJ...745L..29D} {745, L29}

\bibitem[\protect\citeauthoryear{{Dolag}, {Komatsu}  \& {Sunyaev}}{{Dolag}
  et~al.}{2016}]{Dolag2016}
{Dolag} K.,  {Komatsu} E.,   {Sunyaev} R.,  2016, \mn@doi [\mnras]
  {10.1093/mnras/stw2035}, \href
  {http://cdsads.u-strasbg.fr/abs/2016MNRAS.463.1797D} {463, 1797}

\bibitem[\protect\citeauthoryear{{Dong} \& {Stone}}{{Dong} \&
  {Stone}}{2009}]{Dong2009}
{Dong} R.,  {Stone} J.~M.,  2009, \mn@doi [\apj]
  {10.1088/0004-637X/704/2/1309}, \href
  {http://cdsads.u-strasbg.fr/abs/2009ApJ...704.1309D} {704, 1309}

\bibitem[\protect\citeauthoryear{{Dong}, {Rasmussen}  \& {Mulchaey}}{{Dong}
  et~al.}{2010}]{Dong2010}
{Dong} R.,  {Rasmussen} J.,   {Mulchaey} J.~S.,  2010, \mn@doi [\apj]
  {10.1088/0004-637X/712/2/883}, \href
  {http://cdsads.u-strasbg.fr/abs/2010ApJ...712..883D} {712, 883}

\bibitem[\protect\citeauthoryear{{Dubois}, {Peirani}, {Pichon}, {Devriendt},
  {Gavazzi}, {Welker}  \& {Volonteri}}{{Dubois} et~al.}{2016}]{Dubois2016}
{Dubois} Y.,  {Peirani} S.,  {Pichon} C.,  {Devriendt} J.,  {Gavazzi} R.,
  {Welker} C.,   {Volonteri} M.,  2016, \mn@doi [\mnras]
  {10.1093/mnras/stw2265}, \href
  {http://cdsads.u-strasbg.fr/abs/2016MNRAS.463.3948D} {463, 3948}

\bibitem[\protect\citeauthoryear{{Duffell} \& {MacFadyen}}{{Duffell} \&
  {MacFadyen}}{2011}]{Duffell2011}
{Duffell} P.~C.,  {MacFadyen} A.~I.,  2011, \mn@doi [\apjs]
  {10.1088/0067-0049/197/2/15}, \href
  {http://cdsads.u-strasbg.fr/abs/2011ApJS..197...15D} {197, 15}

\bibitem[\protect\citeauthoryear{{Dunn}, {Fabian}  \& {Taylor}}{{Dunn}
  et~al.}{2005}]{Dunn2005}
{Dunn} R.~J.~H.,  {Fabian} A.~C.,   {Taylor} G.~B.,  2005, \mn@doi [\mnras]
  {10.1111/j.1365-2966.2005.09673.x}, \href
  {http://cdsads.u-strasbg.fr/abs/2005MNRAS.364.1343D} {364, 1343}

\bibitem[\protect\citeauthoryear{{Dursi} \& {Pfrommer}}{{Dursi} \&
  {Pfrommer}}{2008}]{Dursi2008}
{Dursi} L.~J.,  {Pfrommer} C.,  2008, \mn@doi [\apj] {10.1086/529371}, \href
  {http://cdsads.u-strasbg.fr/abs/2008ApJ...677..993D} {677, 993}

\bibitem[\protect\citeauthoryear{{Eckert}, {Gaspari}, {Vazza}, {Gastaldello},
  {Tramacere}, {Zimmer}, {Ettori}  \& {Paltani}}{{Eckert}
  et~al.}{2017}]{Eckert2017}
{Eckert} D.,  {Gaspari} M.,  {Vazza} F.,  {Gastaldello} F.,  {Tramacere} A.,
  {Zimmer} S.,  {Ettori} S.,   {Paltani} S.,  2017, \mn@doi [\apjl]
  {10.3847/2041-8213/aa7c1a}, \href
  {http://cdsads.u-strasbg.fr/abs/2017ApJ...843L..29E} {843, L29}

\bibitem[\protect\citeauthoryear{{Evrard}}{{Evrard}}{1997}]{Evrard1997}
{Evrard} A.~E.,  1997, \mn@doi [\mnras] {10.1093/mnras/292.2.289}, \href
  {http://cdsads.u-strasbg.fr/abs/1997MNRAS.292..289E} {292, 289}

\bibitem[\protect\citeauthoryear{{Fabian}}{{Fabian}}{1994}]{Fabian1994}
{Fabian} A.~C.,  1994, \mn@doi [\araa] {10.1146/annurev.aa.32.090194.001425},
  \href {http://cdsads.u-strasbg.fr/abs/1994ARA%26A..32..277F} {32, 277}

\bibitem[\protect\citeauthoryear{{Fabian}}{{Fabian}}{2012}]{Fabian2012}
{Fabian} A.~C.,  2012, \mn@doi [\araa] {10.1146/annurev-astro-081811-125521},
  \href {http://cdsads.u-strasbg.fr/abs/2012ARA%26A..50..455F} {50, 455}

\bibitem[\protect\citeauthoryear{{Fabian} \& {Nulsen}}{{Fabian} \&
  {Nulsen}}{1977}]{Fabian1977}
{Fabian} A.~C.,  {Nulsen} P.~E.~J.,  1977, \mn@doi [\mnras]
  {10.1093/mnras/180.3.479}, \href
  {http://cdsads.u-strasbg.fr/abs/1977MNRAS.180..479F} {180, 479}

\bibitem[\protect\citeauthoryear{{Fabian} et~al.,}{{Fabian}
  et~al.}{2000}]{Fabian2000}
{Fabian} A.~C.,  et~al., 2000, \mn@doi [\mnras]
  {10.1046/j.1365-8711.2000.03904.x}, \href
  {http://cdsads.u-strasbg.fr/abs/2000MNRAS.318L..65F} {318, L65}

\bibitem[\protect\citeauthoryear{{Forman} et~al.,}{{Forman}
  et~al.}{2007}]{Forman2007}
{Forman} W.,  et~al., 2007, \mn@doi [\apj] {10.1086/519480}, \href
  {http://cdsads.u-strasbg.fr/abs/2007ApJ...665.1057F} {665, 1057}

\bibitem[\protect\citeauthoryear{{Frenk} et~al.,}{{Frenk}
  et~al.}{1999}]{Frenk1999}
{Frenk} C.~S.,  et~al., 1999, \mn@doi [\apj] {10.1086/307908}, \href
  {http://cdsads.u-strasbg.fr/abs/1999ApJ...525..554F} {525, 554}

\bibitem[\protect\citeauthoryear{{Gaburov}, {Johansen}  \& {Levin}}{{Gaburov}
  et~al.}{2012}]{Gaburov2012}
{Gaburov} E.,  {Johansen} A.,   {Levin} Y.,  2012, \mn@doi [\apj]
  {10.1088/0004-637X/758/2/103}, \href
  {http://cdsads.u-strasbg.fr/abs/2012ApJ...758..103G} {758, 103}

\bibitem[\protect\citeauthoryear{{Gaspari}, {Melioli}, {Brighenti}  \&
  {D'Ercole}}{{Gaspari} et~al.}{2011}]{Gaspari2011}
{Gaspari} M.,  {Melioli} C.,  {Brighenti} F.,   {D'Ercole} A.,  2011, \mn@doi
  [\mnras] {10.1111/j.1365-2966.2010.17688.x}, \href
  {http://cdsads.u-strasbg.fr/abs/2011MNRAS.411..349G} {411, 349}

\bibitem[\protect\citeauthoryear{{Gingold} \& {Monaghan}}{{Gingold} \&
  {Monaghan}}{1977}]{Gingold1977}
{Gingold} R.~A.,  {Monaghan} J.~J.,  1977, \mn@doi [\mnras]
  {10.1093/mnras/181.3.375}, \href
  {http://cdsads.u-strasbg.fr/abs/1977MNRAS.181..375G} {181, 375}

\bibitem[\protect\citeauthoryear{{Gitti}, {Feretti}  \& {Schindler}}{{Gitti}
  et~al.}{2006}]{Gitti2006}
{Gitti} M.,  {Feretti} L.,   {Schindler} S.,  2006, \mn@doi [\aap]
  {10.1051/0004-6361:20053998}, \href
  {http://cdsads.u-strasbg.fr/abs/2006A%26A...448..853G} {448, 853}

\bibitem[\protect\citeauthoryear{{Godfrey} \& {Shabala}}{{Godfrey} \&
  {Shabala}}{2013}]{Godfrey2013}
{Godfrey} L.~E.~H.,  {Shabala} S.~S.,  2013, \mn@doi [\apj]
  {10.1088/0004-637X/767/1/12}, \href
  {http://cdsads.u-strasbg.fr/abs/2013ApJ...767...12G} {767, 12}

\bibitem[\protect\citeauthoryear{{Guo} \& {Mathews}}{{Guo} \&
  {Mathews}}{2011}]{Guo2011}
{Guo} F.,  {Mathews} W.~G.,  2011, \mn@doi [\apj]
  {10.1088/0004-637X/728/2/121}, \href
  {http://cdsads.u-strasbg.fr/abs/2011ApJ...728..121G} {728, 121}

\bibitem[\protect\citeauthoryear{{Hahn}, {Martizzi}, {Wu}, {Evrard}, {Teyssier}
   \& {Wechsler}}{{Hahn} et~al.}{2017}]{Hahn2017}
{Hahn} O.,  {Martizzi} D.,  {Wu} H.-Y.,  {Evrard} A.~E.,  {Teyssier} R.,
  {Wechsler} R.~H.,  2017, \mn@doi [\mnras] {10.1093/mnras/stx001}, 470, 166

\bibitem[\protect\citeauthoryear{{Hitomi Collaboration} et~al.,}{{Hitomi
  Collaboration} et~al.}{2016}]{Hitomi2016}
{Hitomi Collaboration} et~al., 2016, \mn@doi [\nat] {10.1038/nature18627},
  \href {http://cdsads.u-strasbg.fr/abs/2016Natur.535..117H} {535, 117}

\bibitem[\protect\citeauthoryear{{Hopkins}}{{Hopkins}}{2013}]{Hopkins2013}
{Hopkins} P.~F.,  2013, \mn@doi [\mnras] {10.1093/mnras/sts210}, \href
  {http://cdsads.u-strasbg.fr/abs/2013MNRAS.428.2840H} {428, 2840}

\bibitem[\protect\citeauthoryear{{Hopkins}}{{Hopkins}}{2015}]{Hopkins2015}
{Hopkins} P.~F.,  2015, \mn@doi [\mnras] {10.1093/mnras/stv195}, 450, 53

\bibitem[\protect\citeauthoryear{{Hurier} \& {Angulo}}{{Hurier} \&
  {Angulo}}{2017}]{Hurier2017}
{Hurier} G.,  {Angulo} R.~E.,  2017, preprint, \href
  {http://adsabs.harvard.edu/abs/2017arXiv171106029H} {} (\mn@eprint {arXiv}
  {1711.06029})

\bibitem[\protect\citeauthoryear{{Inutsuka}}{{Inutsuka}}{2002}]{Inutsuka2002}
{Inutsuka} S.-I.,  2002, \mn@doi [Journal of Computational Physics]
  {10.1006/jcph.2002.7053}, \href
  {http://cdsads.u-strasbg.fr/abs/2002JCoPh.179..238I} {179, 238}

\bibitem[\protect\citeauthoryear{{Kawata} \& {Gibson}}{{Kawata} \&
  {Gibson}}{2005}]{Kawata2005}
{Kawata} D.,  {Gibson} B.~K.,  2005, \mn@doi [\mnras]
  {10.1111/j.1745-3933.2005.00018.x}, \href
  {http://cdsads.u-strasbg.fr/abs/2005MNRAS.358L..16K} {358, L16}

\bibitem[\protect\citeauthoryear{{Komatsu} \& {Seljak}}{{Komatsu} \&
  {Seljak}}{2001}]{Komatsu2001}
{Komatsu} E.,  {Seljak} U.,  2001, \mn@doi [\mnras]
  {10.1046/j.1365-8711.2001.04838.x}, 327, 1353

\bibitem[\protect\citeauthoryear{{Komatsu} et~al.,}{{Komatsu}
  et~al.}{2011}]{Komatsu2011}
{Komatsu} E.,  et~al., 2011, \mn@doi [\apjs] {10.1088/0067-0049/192/2/18},
  \href {http://cdsads.u-strasbg.fr/abs/2011ApJS..192...18K} {192, 18}

\bibitem[\protect\citeauthoryear{{Kravtsov} \& {Borgani}}{{Kravtsov} \&
  {Borgani}}{2012}]{Kravtsov2012}
{Kravtsov} A.~V.,  {Borgani} S.,  2012, \mn@doi [\araa]
  {10.1146/annurev-astro-081811-125502}, \href
  {http://cdsads.u-strasbg.fr/abs/2012ARA%26A..50..353K} {50, 353}

\bibitem[\protect\citeauthoryear{{Landau} \& {Lifshitz}}{{Landau} \&
  {Lifshitz}}{1959}]{Landau1959}
{Landau} L.~D.,  {Lifshitz} E.~M.,  1959, {Fluid mechanics}.
Oxford: Pergamon Press

\bibitem[\protect\citeauthoryear{{Lanz}, {Ogle}, {Evans}, {Appleton},
  {Guillard}  \& {Emonts}}{{Lanz} et~al.}{2015}]{Lanz2015}
{Lanz} L.,  {Ogle} P.~M.,  {Evans} D.,  {Appleton} P.~N.,  {Guillard} P.,
  {Emonts} B.,  2015, \mn@doi [\apj] {10.1088/0004-637X/801/1/17}, \href
  {http://cdsads.u-strasbg.fr/abs/2015ApJ...801...17L} {801, 17}

\bibitem[\protect\citeauthoryear{{Lau}, {Kravtsov}  \& {Nagai}}{{Lau}
  et~al.}{2009}]{Lau2009}
{Lau} E.~T.,  {Kravtsov} A.~V.,   {Nagai} D.,  2009, \mn@doi [\apj]
  {10.1088/0004-637X/705/2/1129}, \href
  {http://cdsads.u-strasbg.fr/abs/2009ApJ...705.1129L} {705, 1129}

\bibitem[\protect\citeauthoryear{{Lau}, {Gaspari}, {Nagai}  \& {Coppi}}{{Lau}
  et~al.}{2017}]{Lau2017}
{Lau} E.~T.,  {Gaspari} M.,  {Nagai} D.,   {Coppi} P.,  2017, \mn@doi [\apj]
  {10.3847/1538-4357/aa8c00}, \href
  {http://cdsads.u-strasbg.fr/abs/2017ApJ...849...54L} {849, 54}

\bibitem[\protect\citeauthoryear{{Le Brun}, {McCarthy}, {Schaye}  \&
  {Ponman}}{{Le Brun} et~al.}{2014}]{LeBrun2014}
{Le Brun} A.~M.~C.,  {McCarthy} I.~G.,  {Schaye} J.,   {Ponman} T.~J.,  2014,
  \mn@doi [\mnras] {10.1093/mnras/stu608}, 441, 1270

\bibitem[\protect\citeauthoryear{{Lea}, {Silk}, {Kellogg}  \& {Murray}}{{Lea}
  et~al.}{1973}]{Lea1973}
{Lea} S.~M.,  {Silk} J.,  {Kellogg} E.,   {Murray} S.,  1973, \mn@doi [\apjl]
  {10.1086/181300}, \href {http://cdsads.u-strasbg.fr/abs/1973ApJ...184L.105L}
  {184, L105}

\bibitem[\protect\citeauthoryear{{Lucy}}{{Lucy}}{1977}]{Lucy1977}
{Lucy} L.~B.,  1977, \mn@doi [\aj] {10.1086/112164}, \href
  {http://cdsads.u-strasbg.fr/abs/1977AJ.....82.1013L} {82, 1013}

\bibitem[\protect\citeauthoryear{{Mantz}, {Allen}, {Morris}, {Rapetti},
  {Applegate}, {Kelly}, {von der Linden}  \& {Schmidt}}{{Mantz}
  et~al.}{2014}]{Mantz2014}
{Mantz} A.~B.,  {Allen} S.~W.,  {Morris} R.~G.,  {Rapetti} D.~A.,  {Applegate}
  D.~E.,  {Kelly} P.~L.,  {von der Linden} A.,   {Schmidt} R.~W.,  2014,
  \mn@doi [\mnras] {10.1093/mnras/stu368}, \href
  {http://cdsads.u-strasbg.fr/abs/2014MNRAS.440.2077M} {440, 2077}

\bibitem[\protect\citeauthoryear{{Martizzi}, {Hahn}, {Wu}, {Evrard}, {Teyssier}
   \& {Wechsler}}{{Martizzi} et~al.}{2016}]{Martizzi2016}
{Martizzi} D.,  {Hahn} O.,  {Wu} H.-Y.,  {Evrard} A.~E.,  {Teyssier} R.,
  {Wechsler} R.~H.,  2016, \mn@doi [\mnras] {10.1093/mnras/stw897}, \href
  {http://cdsads.u-strasbg.fr/abs/2016MNRAS.459.4408M} {459, 4408}

\bibitem[\protect\citeauthoryear{{Mathews} \& {Bregman}}{{Mathews} \&
  {Bregman}}{1978}]{Mathews1978}
{Mathews} W.~G.,  {Bregman} J.~N.,  1978, \mn@doi [\apj] {10.1086/156379},
  \href {http://cdsads.u-strasbg.fr/abs/1978ApJ...224..308M} {224, 308}

\bibitem[\protect\citeauthoryear{{McCarthy} et~al.,}{{McCarthy}
  et~al.}{2010}]{McCarthy2010}
{McCarthy} I.~G.,  et~al., 2010, \mn@doi [\mnras]
  {10.1111/j.1365-2966.2010.16750.x}, \href
  {http://cdsads.u-strasbg.fr/abs/2010MNRAS.406..822M} {406, 822}

\bibitem[\protect\citeauthoryear{{McNamara} \& {Nulsen}}{{McNamara} \&
  {Nulsen}}{2007}]{McNamara2007}
{McNamara} B.~R.,  {Nulsen} P.~E.~J.,  2007, \mn@doi [\araa]
  {10.1146/annurev.astro.45.051806.110625}, \href
  {http://cdsads.u-strasbg.fr/abs/2007ARA%26A..45..117M} {45, 117}

\bibitem[\protect\citeauthoryear{{McNamara} et~al.,}{{McNamara}
  et~al.}{2001}]{McNamara2001}
{McNamara} B.~R.,  et~al., 2001, \mn@doi [\apjl] {10.1086/338326}, \href
  {http://cdsads.u-strasbg.fr/abs/2001ApJ...562L.149M} {562, L149}

\bibitem[\protect\citeauthoryear{{Meece}, {Voit}  \& {O'Shea}}{{Meece}
  et~al.}{2017}]{Meece2017}
{Meece} G.~R.,  {Voit} G.~M.,   {O'Shea} B.~W.,  2017, \mn@doi [\apj]
  {10.3847/1538-4357/aa6fb1}, \href
  {http://cdsads.u-strasbg.fr/abs/2017ApJ...841..133M} {841, 133}

\bibitem[\protect\citeauthoryear{{Mingo}, {Hardcastle}, {Croston}, {Evans},
  {Kharb}, {Kraft}  \& {Lenc}}{{Mingo} et~al.}{2012}]{Mingo2012}
{Mingo} B.,  {Hardcastle} M.~J.,  {Croston} J.~H.,  {Evans} D.~A.,  {Kharb} P.,
   {Kraft} R.~P.,   {Lenc} E.,  2012, \mn@doi [\apj]
  {10.1088/0004-637X/758/2/95}, \href
  {http://cdsads.u-strasbg.fr/abs/2012ApJ...758...95M} {758, 95}

\bibitem[\protect\citeauthoryear{{Miniati}}{{Miniati}}{2014}]{Miniati2014}
{Miniati} F.,  2014, \mn@doi [\apj] {10.1088/0004-637X/782/1/21}, \href
  {http://adsabs.harvard.edu/abs/2014ApJ...782...21M} {782, 21}

\bibitem[\protect\citeauthoryear{{Mitchell}, {McCarthy}, {Bower}, {Theuns}  \&
  {Crain}}{{Mitchell} et~al.}{2009}]{Mitchell2009}
{Mitchell} N.~L.,  {McCarthy} I.~G.,  {Bower} R.~G.,  {Theuns} T.,   {Crain}
  R.~A.,  2009, \mn@doi [\mnras] {10.1111/j.1365-2966.2009.14550.x}, \href
  {http://adsabs.harvard.edu/abs/2009MNRAS.395..180M} {395, 180}

\bibitem[\protect\citeauthoryear{{Mo}, {van den Bosch}  \& {White}}{{Mo}
  et~al.}{2010}]{Mo2010}
{Mo} H.,  {van den Bosch} F.~C.,   {White} S.,  2010, {Galaxy Formation and
  Evolution}

\bibitem[\protect\citeauthoryear{{Monaghan}}{{Monaghan}}{1992}]{Monaghan1992}
{Monaghan} J.~J.,  1992, \mn@doi [\araa] {10.1146/annurev.aa.30.090192.002551},
  \href {http://cdsads.u-strasbg.fr/abs/1992ARA%26A..30..543M} {30, 543}

\bibitem[\protect\citeauthoryear{Monaghan}{Monaghan}{2012}]{Monaghan2012}
Monaghan J.,  2012, \mn@doi [Annual Review of Fluid Mechanics]
  {10.1146/annurev-fluid-120710-101220}, 44, 323

\bibitem[\protect\citeauthoryear{{Nagai}, {Kravtsov}  \& {Vikhlinin}}{{Nagai}
  et~al.}{2007}]{Nagai2007}
{Nagai} D.,  {Kravtsov} A.~V.,   {Vikhlinin} A.,  2007, \mn@doi [\apj]
  {10.1086/521328}, \href {http://cdsads.u-strasbg.fr/abs/2007ApJ...668....1N}
  {668, 1}

\bibitem[\protect\citeauthoryear{{Navarro}, {Frenk}  \& {White}}{{Navarro}
  et~al.}{1997}]{Navarro1997}
{Navarro} J.~F.,  {Frenk} C.~S.,   {White} S.~D.~M.,  1997, \mn@doi [\apj]
  {10.1086/304888}, 490, 493

\bibitem[\protect\citeauthoryear{{Nelson}, {Lau}, {Nagai}, {Rudd}  \&
  {Yu}}{{Nelson} et~al.}{2014}]{Nelson2014}
{Nelson} K.,  {Lau} E.~T.,  {Nagai} D.,  {Rudd} D.~H.,   {Yu} L.,  2014,
  \mn@doi [\apj] {10.1088/0004-637X/782/2/107}, \href
  {http://cdsads.u-strasbg.fr/abs/2014ApJ...782..107N} {782, 107}

\bibitem[\protect\citeauthoryear{{Nemmen}, {Georganopoulos}, {Guiriec},
  {Meyer}, {Gehrels}  \& {Sambruna}}{{Nemmen} et~al.}{2012}]{Nemmen2012}
{Nemmen} R.~S.,  {Georganopoulos} M.,  {Guiriec} S.,  {Meyer} E.~T.,  {Gehrels}
  N.,   {Sambruna} R.~M.,  2012, \mn@doi [Science] {10.1126/science.1227416},
  \href {http://cdsads.u-strasbg.fr/abs/2012Sci...338.1445N} {338, 1445}

\bibitem[\protect\citeauthoryear{{Ohto}, {Kawano}  \& {Fukazawa}}{{Ohto}
  et~al.}{2003}]{Ohto2003}
{Ohto} A.,  {Kawano} N.,   {Fukazawa} Y.,  2003, \mn@doi [\pasj]
  {10.1093/pasj/55.4.819}, \href
  {http://cdsads.u-strasbg.fr/abs/2003PASJ...55..819O} {55, 819}

\bibitem[\protect\citeauthoryear{{Okamoto}, {Jenkins}, {Eke}, {Quilis}  \&
  {Frenk}}{{Okamoto} et~al.}{2003}]{Okamoto2003}
{Okamoto} T.,  {Jenkins} A.,  {Eke} V.~R.,  {Quilis} V.,   {Frenk} C.~S.,
  2003, \mn@doi [\mnras] {10.1046/j.1365-8711.2003.06948.x}, \href
  {http://cdsads.u-strasbg.fr/abs/2003MNRAS.345..429O} {345, 429}

\bibitem[\protect\citeauthoryear{{Okamoto}, {Nemmen}  \& {Bower}}{{Okamoto}
  et~al.}{2008}]{Okamoto2008}
{Okamoto} T.,  {Nemmen} R.~S.,   {Bower} R.~G.,  2008, \mn@doi [\mnras]
  {10.1111/j.1365-2966.2008.12883.x}, \href
  {http://cdsads.u-strasbg.fr/abs/2008MNRAS.385..161O} {385, 161}

\bibitem[\protect\citeauthoryear{{Panagoulia}, {Fabian}, {Sanders}  \&
  {Hlavacek-Larrondo}}{{Panagoulia} et~al.}{2014}]{Panagoulia2014}
{Panagoulia} E.~K.,  {Fabian} A.~C.,  {Sanders} J.~S.,   {Hlavacek-Larrondo}
  J.,  2014, \mn@doi [\mnras] {10.1093/mnras/stu1499}, \href
  {http://cdsads.u-strasbg.fr/abs/2014MNRAS.444.1236P} {444, 1236}

\bibitem[\protect\citeauthoryear{{Perucho}, {Mart{\'{\i}}}, {Quilis}  \&
  {Ricciardelli}}{{Perucho} et~al.}{2014}]{Perucho2014}
{Perucho} M.,  {Mart{\'{\i}}} J.-M.,  {Quilis} V.,   {Ricciardelli} E.,  2014,
  \mn@doi [\mnras] {10.1093/mnras/stu1828}, \href
  {http://cdsads.u-strasbg.fr/abs/2014MNRAS.445.1462P} {445, 1462}

\bibitem[\protect\citeauthoryear{{Planck Collaboration} et~al.,}{{Planck
  Collaboration} et~al.}{2016}]{Planck2016a}
{Planck Collaboration} et~al., 2016, \mn@doi [\aap]
  {10.1051/0004-6361/201525830}, \href
  {http://adsabs.harvard.edu/abs/2016A%26A...594A..13P} {594, A13}

\bibitem[\protect\citeauthoryear{{Planelles}, {Borgani}, {Fabjan}, {Killedar},
  {Murante}, {Granato}, {Ragone-Figueroa}  \& {Dolag}}{{Planelles}
  et~al.}{2014}]{Planelles2014}
{Planelles} S.,  {Borgani} S.,  {Fabjan} D.,  {Killedar} M.,  {Murante} G.,
  {Granato} G.~L.,  {Ragone-Figueroa} C.,   {Dolag} K.,  2014, \mn@doi [\mnras]
  {10.1093/mnras/stt2141}, \href
  {http://cdsads.u-strasbg.fr/abs/2014MNRAS.438..195P} {438, 195}

\bibitem[\protect\citeauthoryear{{Pontzen}, {Ro{\v s}kar}, {Stinson}, {Woods},
  {Reed}, {Coles}  \& {Quinn}}{{Pontzen} et~al.}{2013}]{pynbody}
{Pontzen} A.,  {Ro{\v s}kar} R.,  {Stinson} G.~S.,  {Woods} R.,  {Reed} D.~M.,
  {Coles} J.,   {Quinn} T.~R.,  2013, {pynbody: Astrophysics Simulation
  Analysis for Python}

\bibitem[\protect\citeauthoryear{{Power}, {Read}  \& {Hobbs}}{{Power}
  et~al.}{2014}]{Power2014}
{Power} C.,  {Read} J.~I.,   {Hobbs} A.,  2014, \mn@doi [\mnras]
  {10.1093/mnras/stu418}, \href
  {http://adsabs.harvard.edu/abs/2014MNRAS.440.3243P} {440, 3243}

\bibitem[\protect\citeauthoryear{{Pratt} et~al.,}{{Pratt}
  et~al.}{2010}]{Pratt2010}
{Pratt} G.~W.,  et~al., 2010, \mn@doi [\aap] {10.1051/0004-6361/200913309},
  \href {http://cdsads.u-strasbg.fr/abs/2010A%26A...511A..85P} {511, A85}

\bibitem[\protect\citeauthoryear{{Price}}{{Price}}{2012}]{Price2012}
{Price} D.~J.,  2012, \mn@doi [Journal of Computational Physics]
  {10.1016/j.jcp.2010.12.011}, \href
  {http://cdsads.u-strasbg.fr/abs/2012JCoPh.231..759P} {231, 759}

\bibitem[\protect\citeauthoryear{{Rasia} et~al.,}{{Rasia}
  et~al.}{2015}]{Rasia2015}
{Rasia} E.,  et~al., 2015, \mn@doi [\apjl] {10.1088/2041-8205/813/1/L17}, \href
  {http://cdsads.u-strasbg.fr/abs/2015ApJ...813L..17R} {813, L17}

\bibitem[\protect\citeauthoryear{{Read}, {Hayfield}  \& {Agertz}}{{Read}
  et~al.}{2010}]{Read2010}
{Read} J.~I.,  {Hayfield} T.,   {Agertz} O.,  2010, \mn@doi [\mnras]
  {10.1111/j.1365-2966.2010.16577.x}, \href
  {http://cdsads.u-strasbg.fr/abs/2010MNRAS.405.1513R} {405, 1513}

\bibitem[\protect\citeauthoryear{{Reynolds}, {McKernan}, {Fabian}, {Stone}  \&
  {Vernaleo}}{{Reynolds} et~al.}{2005}]{Reynolds2005}
{Reynolds} C.~S.,  {McKernan} B.,  {Fabian} A.~C.,  {Stone} J.~M.,   {Vernaleo}
  J.~C.,  2005, \mn@doi [\mnras] {10.1111/j.1365-2966.2005.08643.x}, \href
  {http://cdsads.u-strasbg.fr/abs/2005MNRAS.357..242R} {357, 242}

\bibitem[\protect\citeauthoryear{{Reynolds}, {Balbus}  \&
  {Schekochihin}}{{Reynolds} et~al.}{2015}]{Reynolds2015}
{Reynolds} C.~S.,  {Balbus} S.~A.,   {Schekochihin} A.~A.,  2015, \mn@doi
  [\apj] {10.1088/0004-637X/815/1/41}, \href
  {http://cdsads.u-strasbg.fr/abs/2015ApJ...815...41R} {815, 41}

\bibitem[\protect\citeauthoryear{{Ritchie} \& {Thomas}}{{Ritchie} \&
  {Thomas}}{2001}]{Ritchie2001}
{Ritchie} B.~W.,  {Thomas} P.~A.,  2001, \mn@doi [\mnras]
  {10.1046/j.1365-8711.2001.04268.x}, \href
  {http://cdsads.u-strasbg.fr/abs/2001MNRAS.323..743R} {323, 743}

\bibitem[\protect\citeauthoryear{{Robinson} et~al.,}{{Robinson}
  et~al.}{2004}]{Robinson2004}
{Robinson} K.,  et~al., 2004, \mn@doi [\apj] {10.1086/380817}, \href
  {http://cdsads.u-strasbg.fr/abs/2004ApJ...601..621R} {601, 621}

\bibitem[\protect\citeauthoryear{{Roediger}, {Br{\"u}ggen}, {Rebusco},
  {B{\"o}hringer}  \& {Churazov}}{{Roediger} et~al.}{2007}]{Roediger2007}
{Roediger} E.,  {Br{\"u}ggen} M.,  {Rebusco} P.,  {B{\"o}hringer} H.,
  {Churazov} E.,  2007, \mn@doi [\mnras] {10.1111/j.1365-2966.2006.11300.x},
  \href {http://cdsads.u-strasbg.fr/abs/2007MNRAS.375...15R} {375, 15}

\bibitem[\protect\citeauthoryear{{Rosswog}}{{Rosswog}}{2009}]{Rosswog2009}
{Rosswog} S.,  2009, \mn@doi [\nar] {10.1016/j.newar.2009.08.007}, \href
  {http://cdsads.u-strasbg.fr/abs/2009NewAR..53...78R} {53, 78}

\bibitem[\protect\citeauthoryear{{Saitoh} \& {Makino}}{{Saitoh} \&
  {Makino}}{2013}]{Saitoh2013}
{Saitoh} T.~R.,  {Makino} J.,  2013, \mn@doi [\apj]
  {10.1088/0004-637X/768/1/44}, \href
  {http://cdsads.u-strasbg.fr/abs/2013ApJ...768...44S} {768, 44}

\bibitem[\protect\citeauthoryear{{Sanders}, {Fabian}  \& {Taylor}}{{Sanders}
  et~al.}{2009}]{Sanders2009}
{Sanders} J.~S.,  {Fabian} A.~C.,   {Taylor} G.~B.,  2009, \mn@doi [\mnras]
  {10.1111/j.1365-2966.2008.14207.x}, \href
  {http://cdsads.u-strasbg.fr/abs/2009MNRAS.393...71S} {393, 71}

\bibitem[\protect\citeauthoryear{{Sanderson}, {Finoguenov}  \&
  {Mohr}}{{Sanderson} et~al.}{2005}]{Sanderson2005}
{Sanderson} A.~J.~R.,  {Finoguenov} A.,   {Mohr} J.~J.,  2005, \mn@doi [\apj]
  {10.1086/431750}, \href {http://cdsads.u-strasbg.fr/abs/2005ApJ...630..191S}
  {630, 191}

\bibitem[\protect\citeauthoryear{{Schaller}, {Dalla Vecchia}, {Schaye},
  {Bower}, {Theuns}, {Crain}, {Furlong}  \& {McCarthy}}{{Schaller}
  et~al.}{2015}]{Schaller2015}
{Schaller} M.,  {Dalla Vecchia} C.,  {Schaye} J.,  {Bower} R.~G.,  {Theuns} T.,
   {Crain} R.~A.,  {Furlong} M.,   {McCarthy} I.~G.,  2015, \mn@doi [\mnras]
  {10.1093/mnras/stv2169}, \href
  {http://cdsads.u-strasbg.fr/abs/2015MNRAS.454.2277S} {454, 2277}

\bibitem[\protect\citeauthoryear{{Schuecker}, {Finoguenov}, {Miniati},
  {B{\"o}hringer}  \& {Briel}}{{Schuecker} et~al.}{2004}]{Schuecker2004}
{Schuecker} P.,  {Finoguenov} A.,  {Miniati} F.,  {B{\"o}hringer} H.,   {Briel}
  U.~G.,  2004, \mn@doi [\aap] {10.1051/0004-6361:20041039}, \href
  {http://cdsads.u-strasbg.fr/abs/2004A%26A...426..387S} {426, 387}

\bibitem[\protect\citeauthoryear{Schwarzschild}{Schwarzschild}{1906}]{Schwarzschild1906}
Schwarzschild K.,  1906, Nachrichten von der Gesellschaft der Wissenschaften zu
  G\"ottingen, Mathematisch-Physikalische Klasse, 1906, 41

\bibitem[\protect\citeauthoryear{{Sembolini} et~al.,}{{Sembolini}
  et~al.}{2016}]{Sembolini2016}
{Sembolini} F.,  et~al., 2016, \mn@doi [\mnras] {10.1093/mnras/stw800}, \href
  {http://adsabs.harvard.edu/abs/2016MNRAS.459.2973S} {459, 2973}

\bibitem[\protect\citeauthoryear{{Shurkin}, {Dunn}, {Gentile}, {Taylor}  \&
  {Allen}}{{Shurkin} et~al.}{2008}]{Shurkin2008}
{Shurkin} K.,  {Dunn} R.~J.~H.,  {Gentile} G.,  {Taylor} G.~B.,   {Allen}
  S.~W.,  2008, \mn@doi [\mnras] {10.1111/j.1365-2966.2007.12651.x}, \href
  {http://cdsads.u-strasbg.fr/abs/2008MNRAS.383..923S} {383, 923}

\bibitem[\protect\citeauthoryear{{Sijacki} \& {Springel}}{{Sijacki} \&
  {Springel}}{2006}]{Sijacki2006}
{Sijacki} D.,  {Springel} V.,  2006, \mn@doi [\mnras]
  {10.1111/j.1365-2966.2005.09860.x}, \href
  {http://cdsads.u-strasbg.fr/abs/2006MNRAS.366..397S} {366, 397}

\bibitem[\protect\citeauthoryear{{Sijacki}, {Springel}, {Di Matteo}  \&
  {Hernquist}}{{Sijacki} et~al.}{2007}]{Sijacki2007}
{Sijacki} D.,  {Springel} V.,  {Di Matteo} T.,   {Hernquist} L.,  2007, \mn@doi
  [\mnras] {10.1111/j.1365-2966.2007.12153.x}, \href
  {http://cdsads.u-strasbg.fr/abs/2007MNRAS.380..877S} {380, 877}

\bibitem[\protect\citeauthoryear{{Springel}}{{Springel}}{2010a}]{Springel2010_SPH}
{Springel} V.,  2010a, \mn@doi [\araa] {10.1146/annurev-astro-081309-130914},
  \href {http://cdsads.u-strasbg.fr/abs/2010ARA%26A..48..391S} {48, 391}

\bibitem[\protect\citeauthoryear{{Springel}}{{Springel}}{2010b}]{Springel2010_Arepo}
{Springel} V.,  2010b, \mn@doi [\mnras] {10.1111/j.1365-2966.2009.15715.x},
  \href {http://cdsads.u-strasbg.fr/abs/2010MNRAS.401..791S} {401, 791}

\bibitem[\protect\citeauthoryear{{Steinborn}, {Dolag}, {Hirschmann}, {Prieto}
  \& {Remus}}{{Steinborn} et~al.}{2015}]{Steinborn2015}
{Steinborn} L.~K.,  {Dolag} K.,  {Hirschmann} M.,  {Prieto} M.~A.,   {Remus}
  R.-S.,  2015, \mn@doi [\mnras] {10.1093/mnras/stv072}, \href
  {http://cdsads.u-strasbg.fr/abs/2015MNRAS.448.1504S} {448, 1504}

\bibitem[\protect\citeauthoryear{{Sutherland} \& {Bicknell}}{{Sutherland} \&
  {Bicknell}}{2007}]{Sutherland2007}
{Sutherland} R.~S.,  {Bicknell} G.~V.,  2007, \mn@doi [\apjs] {10.1086/520640},
  \href {http://cdsads.u-strasbg.fr/abs/2007ApJS..173...37S} {173, 37}

\bibitem[\protect\citeauthoryear{Teyssier}{Teyssier}{2002}]{Teyssier2002}
Teyssier R.,  2002, \mn@doi [\aap] {10.1051/0004-6361}, 385, 337

\bibitem[\protect\citeauthoryear{{Teyssier}, {Moore}, {Martizzi}, {Dubois}  \&
  {Mayer}}{{Teyssier} et~al.}{2011}]{Teyssier2011}
{Teyssier} R.,  {Moore} B.,  {Martizzi} D.,  {Dubois} Y.,   {Mayer} L.,  2011,
  \mn@doi [\mnras] {10.1111/j.1365-2966.2011.18399.x}, \href
  {http://cdsads.u-strasbg.fr/abs/2011MNRAS.414..195T} {414, 195}

\bibitem[\protect\citeauthoryear{{Vazza}, {Gheller}  \& {Brunetti}}{{Vazza}
  et~al.}{2010}]{Vazza2010}
{Vazza} F.,  {Gheller} C.,   {Brunetti} G.,  2010, \mn@doi [\aap]
  {10.1051/0004-6361/200913464}, \href
  {http://cdsads.u-strasbg.fr/abs/2010A%26A...513A..32V} {513, A32}

\bibitem[\protect\citeauthoryear{{Vernaleo} \& {Reynolds}}{{Vernaleo} \&
  {Reynolds}}{2006}]{Vernaleo2006}
{Vernaleo} J.~C.,  {Reynolds} C.~S.,  2006, \mn@doi [\apj] {10.1086/504029},
  \href {http://cdsads.u-strasbg.fr/abs/2006ApJ...645...83V} {645, 83}

\bibitem[\protect\citeauthoryear{{Vogelsberger}, {Genel}, {Sijacki}, {Torrey},
  {Springel}  \& {Hernquist}}{{Vogelsberger} et~al.}{2013}]{Vogelsberger2013}
{Vogelsberger} M.,  {Genel} S.,  {Sijacki} D.,  {Torrey} P.,  {Springel} V.,
  {Hernquist} L.,  2013, \mn@doi [\mnras] {10.1093/mnras/stt1789}, \href
  {http://cdsads.u-strasbg.fr/abs/2013MNRAS.436.3031V} {436, 3031}

\bibitem[\protect\citeauthoryear{{Voit}, {Meece}, {Li}, {O'Shea}, {Bryan}  \&
  {Donahue}}{{Voit} et~al.}{2017}]{Voit2017}
{Voit} G.~M.,  {Meece} G.,  {Li} Y.,  {O'Shea} B.~W.,  {Bryan} G.~L.,
  {Donahue} M.,  2017, \mn@doi [\apj] {10.3847/1538-4357/aa7d04}, \href
  {http://cdsads.u-strasbg.fr/abs/2017ApJ...845...80V} {845, 80}

\bibitem[\protect\citeauthoryear{{Wadsley}, {Veeravalli}  \&
  {Couchman}}{{Wadsley} et~al.}{2008}]{Wadsley2008}
{Wadsley} J.~W.,  {Veeravalli} G.,   {Couchman} H.~M.~P.,  2008, \mn@doi
  [\mnras] {10.1111/j.1365-2966.2008.13260.x}, \href
  {http://cdsads.u-strasbg.fr/abs/2008MNRAS.387..427W} {387, 427}

\bibitem[\protect\citeauthoryear{{Wagner}, {Bicknell}  \& {Umemura}}{{Wagner}
  et~al.}{2012}]{Wagner2012}
{Wagner} A.~Y.,  {Bicknell} G.~V.,   {Umemura} M.,  2012, \mn@doi [\apj]
  {10.1088/0004-637X/757/2/136}, \href
  {http://cdsads.u-strasbg.fr/abs/2012ApJ...757..136W} {757, 136}

\bibitem[\protect\citeauthoryear{{Wiersma}, {Schaye}  \& {Theuns}}{{Wiersma}
  et~al.}{2011}]{Wiersma2011}
{Wiersma} R.~P.~C.,  {Schaye} J.,   {Theuns} T.,  2011, \mn@doi [\mnras]
  {10.1111/j.1365-2966.2011.18709.x}, \href
  {http://cdsads.u-strasbg.fr/abs/2011MNRAS.415..353W} {415, 353}

\bibitem[\protect\citeauthoryear{{Woo} \& {Urry}}{{Woo} \&
  {Urry}}{2002}]{Woo2002}
{Woo} J.-H.,  {Urry} C.~M.,  2002, \mn@doi [\apj] {10.1086/342878}, \href
  {http://cdsads.u-strasbg.fr/abs/2002ApJ...579..530W} {579, 530}

\bibitem[\protect\citeauthoryear{{Zhuravleva} et~al.,}{{Zhuravleva}
  et~al.}{2014}]{Zhuravleva2014}
{Zhuravleva} I.,  et~al., 2014, \mn@doi [\nat] {10.1038/nature13830}, \href
  {http://cdsads.u-strasbg.fr/abs/2014Natur.515...85Z} {515, 85}

\bibitem[\protect\citeauthoryear{{ZuHone}, {Miller}, {Bulbul}  \&
  {Zhuravleva}}{{ZuHone} et~al.}{2017}]{ZuHone2017}
{ZuHone} J.,  {Miller} E.~D.,  {Bulbul} E.,   {Zhuravleva} I.,  2017, preprint,
  \href {http://cdsads.u-strasbg.fr/abs/2017arXiv170807206Z} {} (\mn@eprint
  {arXiv} {1708.07206})

\bibitem[\protect\citeauthoryear{{van Leer}}{{van Leer}}{1979}]{VanLeer1979}
{van Leer} B.,  1979, \mn@doi [Journal of Computational Physics]
  {10.1016/0021-9991(79)90145-1}, \href
  {http://adsabs.harvard.edu/abs/1979JCoPh..32..101V} {32, 101}

\makeatother
\end{thebibliography}

\appendix

\section{A Spherically symmetric 1D MUSCL solver}
\label{sec:solver1d}
In this appendix, we briefly discuss our implementation of a one-dimensional, spherically symmetric hydrodynamics solver which we use to validate the results obtained with the three-dimensional simulation codes.

\subsection{Governing equations and numerical implementation}
The equations of ideal hydrodynamics in conservative form and under spherical symmetry take the form
\begin{eqnarray}
\frac{\partial \rho}{\partial t} + \frac{\partial\left(S\rho u\right)}{\partial V} & = & 0 \\
\frac{\partial \rho u}{\partial t} + \frac{\partial\left(S (\rho u^2 +P)\right)}{\partial V} & = & -\frac{2P}{r} - \frac{\partial \phi}{\partial r}\\
\frac{\partial \rho e\sub{tot}}{\partial t} + \frac{\partial\left(S\rho h\sub{tot} u\right)}{\partial V} & = & 0,
\end{eqnarray}
where $\rho$ is the density, $u$ is the radial fluid velocity, $e\sub{tot}$ the total specific energy, $h\sub{tot}$ the total specific enthalpy, $\phi$ the gravitational potential, and $S$ is the surface area, i.e. $S=4\pi r^2$, at radius $r$. These equations are straightforwardly implemented into a standard 1D MUSCL solver \citep{VanLeer1979}, where only the geometric factors need to be inserted along with the source term $2P/r$. For time integration, we use a second order predictor-corrector scheme, just as \ramses, where the predicted step is calculated using the primitive equations, then an approximate Riemann solver (we use HLL) is called to compute the Riemann fluxes at the cell interfaces. In an equivalent way, the source terms are advanced by a half time step using the old, and by another half step using the new solution after updating with fluxes.

\subsection{Validation of the 1D solver} \label{app:1d_solver}
\begin{figure}
\begin{center}
\includegraphics[width=1\columnwidth]{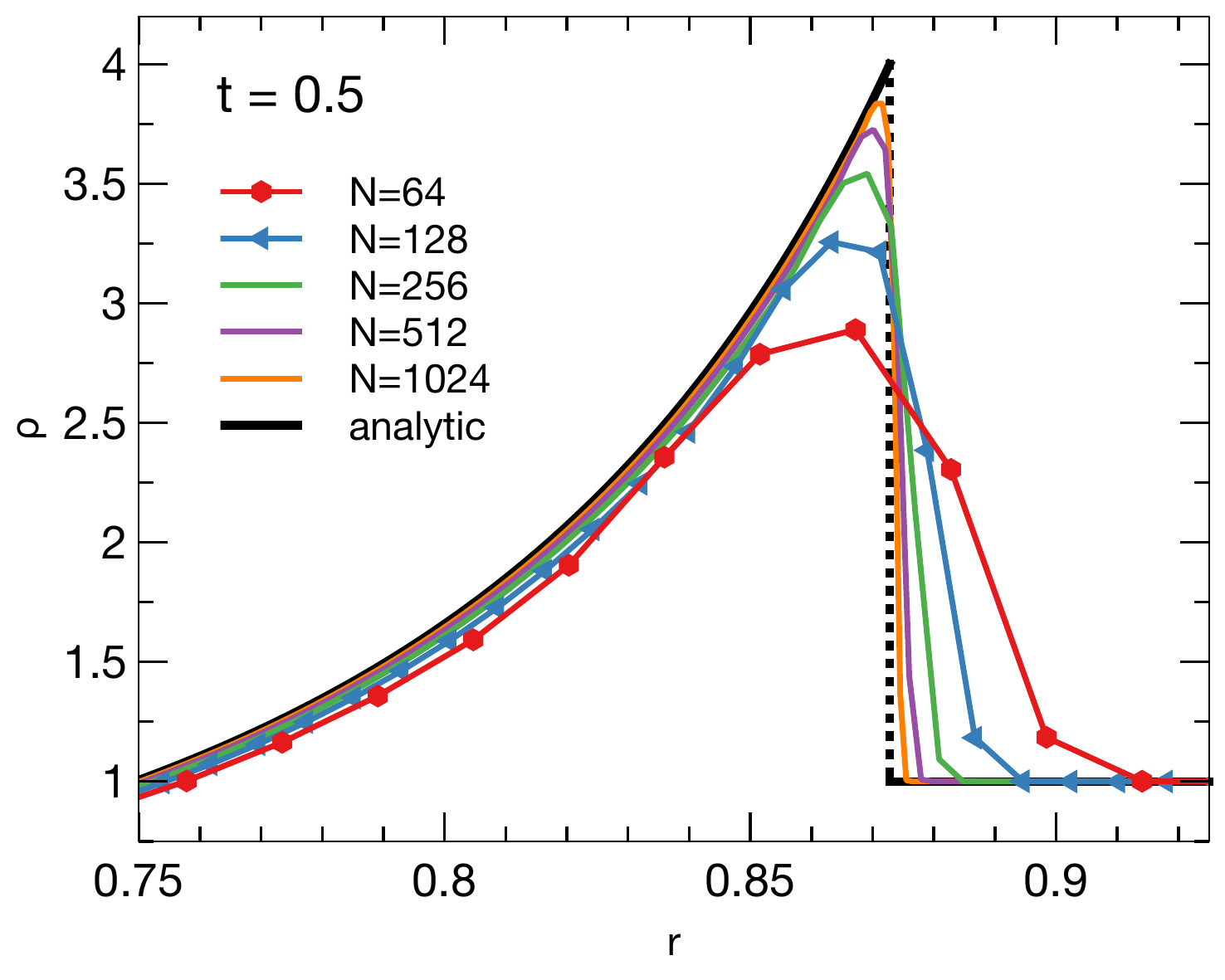}
\end{center}
\caption{\label{fig:sedov_1d_convergence}Convergence of the numerical solution to the Sedov-Taylor blast wave problem with increasing resolution for $N=64, 128, 256, 512$ and $1024$ grid points using our spherical 1D MUSCL scheme. We show results for the dimensionless density $\rho$ at $t=0.5$ for a region around the shock front (which is located at $r\simeq 0.873$).}
\end{figure}

We validate our simple one dimensional solver using the Sedov-Taylor point explosion problem, for which the self-similar analytic solution is known \citep[e.g.][]{Landau1959}. The numerical solution, in dimensionless units, at time $t=0.5$ for an explosion with initial energy $E=1$, expanding into a background of density $\rho=1$, and assuming a polytropic equation of state with exponent $\gamma=5/3$, is shown for various resolutions in comparison to the analytic result in \autoref{fig:sedov_1d_convergence}. We adopted a simulation domain of unit radius, so that the finite volume shells at a resolution of $N$ points have a thickness of $\Delta r= N^{-1}$. The initial energy is inserted in the single innermost shell. We note that at all resolutions the shock has a thickness of $\sim2$ shells.

\section{Density slices from the runs of a central bubble} \label{app:images_bub_cen}

\begin{figure}
\begin{center}
\includegraphics[width=0.5\textwidth]{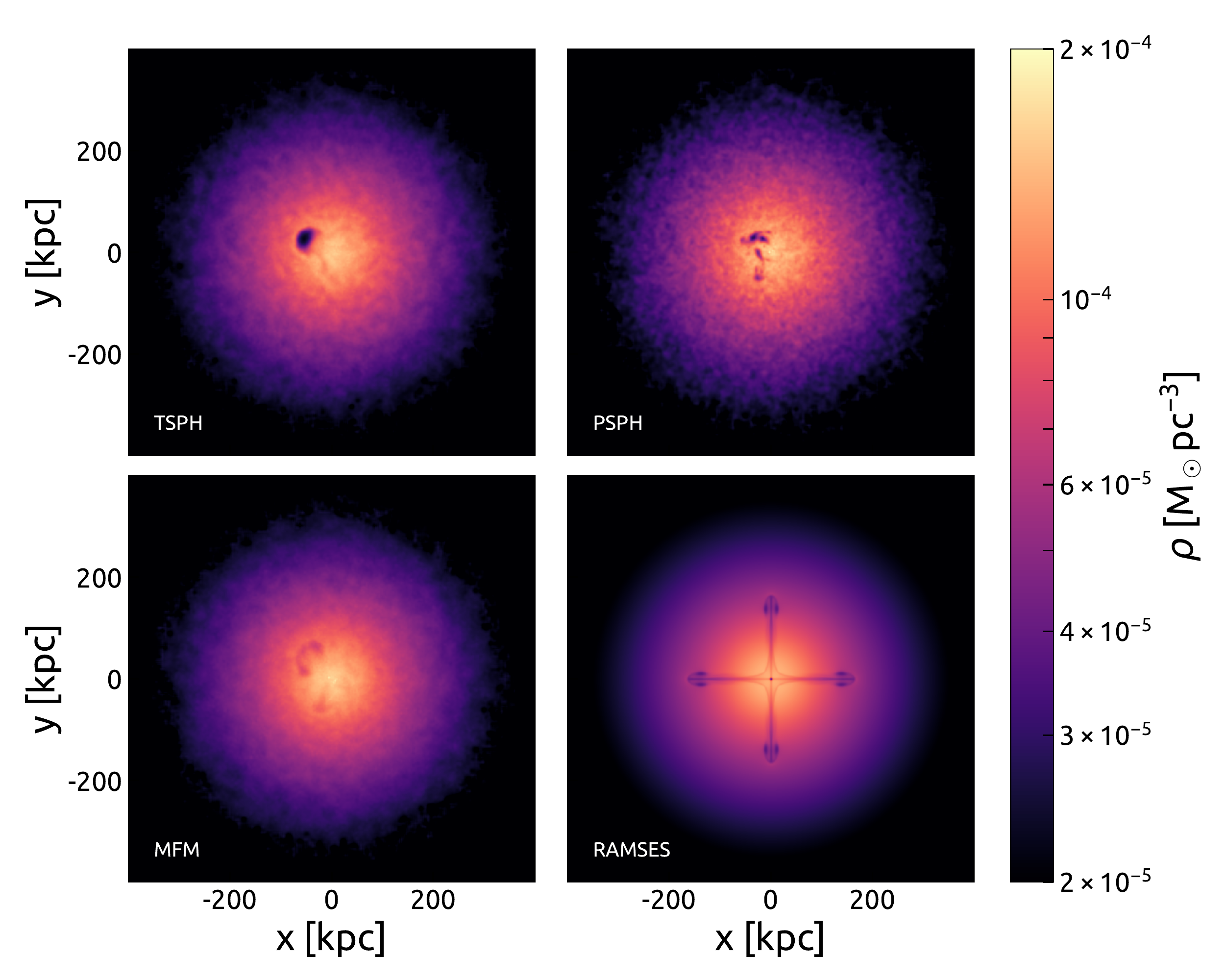}
\end{center}
\caption{
Slices through density distribution for TSPH (top left), PSPH (top right), MFM (bottom left) and RAMSES (bottom right) after 500 Myr of evolution with the central bubble. Each simulation is initialised in the same way - density and temperature follow \citetalias{Komatsu2001} and we artificially raise the temperature of fluid elements in the central sphere with a radius of 10\,kpc to $10^9$\,K.
\label{fig:bub_cen_500}}
\end{figure}

In this appendix, we present the slices of gas density in the runs with the central bubble. \autoref{fig:bub_cen_500} shows the snapshot at $t=500$\,Myr. The bubble rises due to the Poisson noise in the initial condition of the simulations using \gizmo; this breaks the symmetry of the system. The bubble is very pronounced in the TSPH run (upper left) because the growth of the KHI is suppressed by the spurious tension on the bubble surface. In the PSPH (upper right) and MFM (lower left) runs, the bubble is dissolved the KHI. Because of the symmetry and absence of the Poisson noise a cross structure is formed in the \ramses\, run (lower right). 

\section{Resolution effects}\label{app:resoution}
\begin{figure*}
\includegraphics[width=1\textwidth]{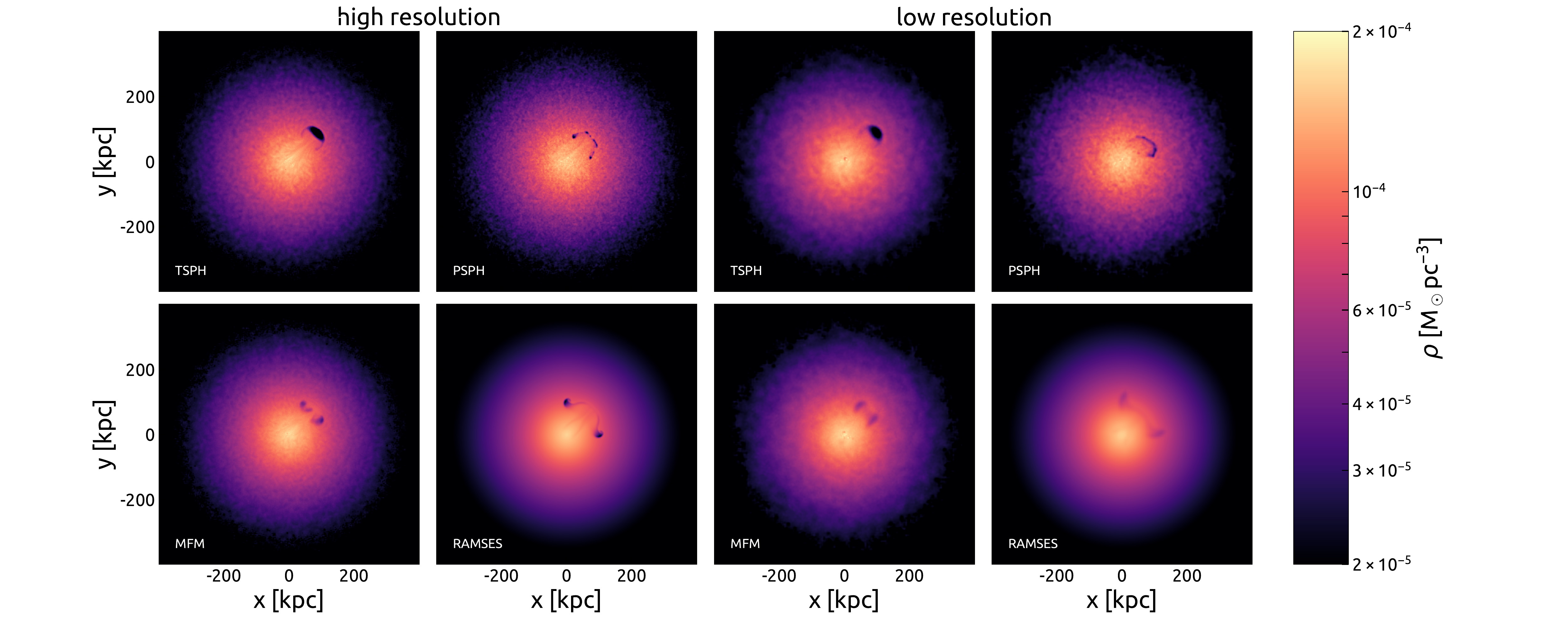}
\caption{
Resolution test. The left four panels show the density slices at $t=500$\,Myr obtained from simulations with the high resolution we adopt as the standard level (64 million particles in the \gizmo\, runs and \texttt{levelmax}=11 in the \ramses\, run) and the right four show those in runs with a degraded resolution (8 million particles in the \gizmo\, runs and \texttt{levelmax}=9 in the \ramses\, run). A shifted hot bubble is initially set and the turbulent velocity field is {\it not} introduced in all runs (for details see \autoref{ssec:shifted_bubble_wo_turbulence}). In each group of panels, the results of simulations using TSPH, PSPH, MFM and RAMSES are illustrated in the upper left, upper right, lower left and lower right panels, respectively. 
\label{fig:bub_10kpc_500_restest}}
\end{figure*}

We study how the results depend on the resolution of simulations. \autoref{fig:bub_10kpc_500_restest} compares simulation results varying the resolution. The left four panels are the density slices from simulations with the high resolution we adopt as the standard level and the right four are those from the runs with a degraded resolution level. The \ramses\, simulations have similar numbers of leaf cells to the number of particles in the \gizmo\, simulations in each resolution level. The overall structures, e.g. direction to which the bubble is rising and position of the bubble, are well captured in the both resolution levels. However, one would find the differences between the two levels of resolution on the small scale, e.g. number of fragmented smaller bubbles and separation between them. In spite of the resolution dependence, our conclusion is not changed because the differences among the hydrodynamical solvers are more significant. 

\bsp	
\label{lastpage}
\end{document}